\documentclass[prb,twocolumn,floatfix,amsmath,amssymb,superscriptaddress]{revtex4}
\pdfoutput=1
\usepackage{latexsym}
\usepackage{graphicx}
\usepackage{times}
\usepackage{amsmath}
\usepackage{dcolumn}
\usepackage{latexsym,amsmath,amssymb,bm,euscript}
\def\be{\begin{equation}}
\def\ee{\end{equation}}
\def\bea{\begin{eqnarray}}
\def\eea{\end{eqnarray}}
\begin{document}

\title{Unusual ordered phases of highly frustrated magnets: a review}

\author{Oleg A. Starykh}
\affiliation{Department of Physics and Astronomy, University of Utah, Salt Lake  City, UT 84112-0830}

\date{December 29, 2014}

\begin{abstract}
We review ground states and excitations of a quantum antiferromagnet on triangular and other frustrated lattices. We pay special attention to 
the combined effects of magnetic field $h$, spatial anisotropy $R$, and spin magnitude $S$. The focus of the review is on the
novel collinear spin density wave and spin nematic states, which are characterized by {\em fully gapped} transverse spin excitations with $S^z = \pm 1$.
We discuss extensively $R-h$ phase diagram of the antiferromagnet, both in the large-$S$ semiclassical limit and the quantum $S=1/2$ limit.
When possible, we point out connections with experimental findings. 

[This is the originally submitted version of the invited review, to be published in Reports on Progress in Physics in 2015. The link, via DOI, to the accepted and published version of the manuscript, which is updated according to the refereesÕ comments, will be provided upon its actual publication.]
\end{abstract}
\maketitle

\tableofcontents
%%%%%%%%%%%%%%%%%%%%%%%%%%%%%%%%%%%%%%%%%%%%%%%%%%%%%%%%%%%%%%%%%%%%%%

\section{Introduction} 
\label{sec:intro}

Frustrated quantum antiferromagnets have been at the center of intense experimental and theoretical investigations
for many years. These relentless efforts have very recently resulted in a number of theoretical and experimental
breakthroughs: quantum entanglement \cite{Zhang2012,Jiang2012a,Grover2013}, 
density matrix renormalization group (DMRG) revolution and Z$_2$ liquids in kagom\'e and $J_1 - J_2$ square lattice models \cite{Yan2011,Jiang2012,Depenbrock2012},
spin-liquid-like behavior in organic Mott insulators \cite{Shimizu2003,Kanoda2011} and kagom\'e lattice antiferromagnet herbertsmithite \cite{Han2012}.

Along the way, a large number of frustrated insulating magnetic materials featuring rather unusual {\em ordered} phases,
such as magnetization plateaux, longitudinal spin-density waves, and spin nematics,
has been discovered and studied. It is these ordered, yet sufficiently unconventional, states of magnetic matter and theoretical
models motivated by them that are the subject of this Key Issue article.

%Most of this 
This review focuses on materials and models based on simple triangular lattice, which, despite many years of fruitful research,
continue to supply us with novel quantum states and phenomena.
Triangular lattice represents, perhaps, the most widely  studied frustrated geometry \cite{Collins1997,Ramirez2001,Balents2010}. 
Indeed, the Ising antiferromagnet on the triangular lattice was the first spin model found to possess a
disordered ground state and extensive residual entropy \cite{Wannier1950} at
zero temperature. While the classical Heisenberg model
on the triangular lattice does order at $T = 0$ into a well-known $120^\circ$
commensurate spiral pattern (also known as a three-sublattice or $\sqrt{3} \times \sqrt{3}$ state), the fate of the quantum 
spin-1/2 Heisenberg Hamiltonian has been the subject of a long and fruitful debate spanning over 30 years of research. 
Eventually it was firmly established that the quantum spin-$1/2$ model remains ordered in the 
classical $120^\circ$ pattern \cite{huse1988simple,Bernu1992,White2007}.
Although the originally proposed resonating valence bond liquid \cite{anderson1973resonating,fazekas1974ground} did not emerge as 
the ground state of the spin-1/2 
Heisenberg model, such a phase was later found in a related quantum dimer model on the triangular lattice.\cite{Moessner2001}

\begin{figure}[h]
\centering
\includegraphics[width=4.0cm]{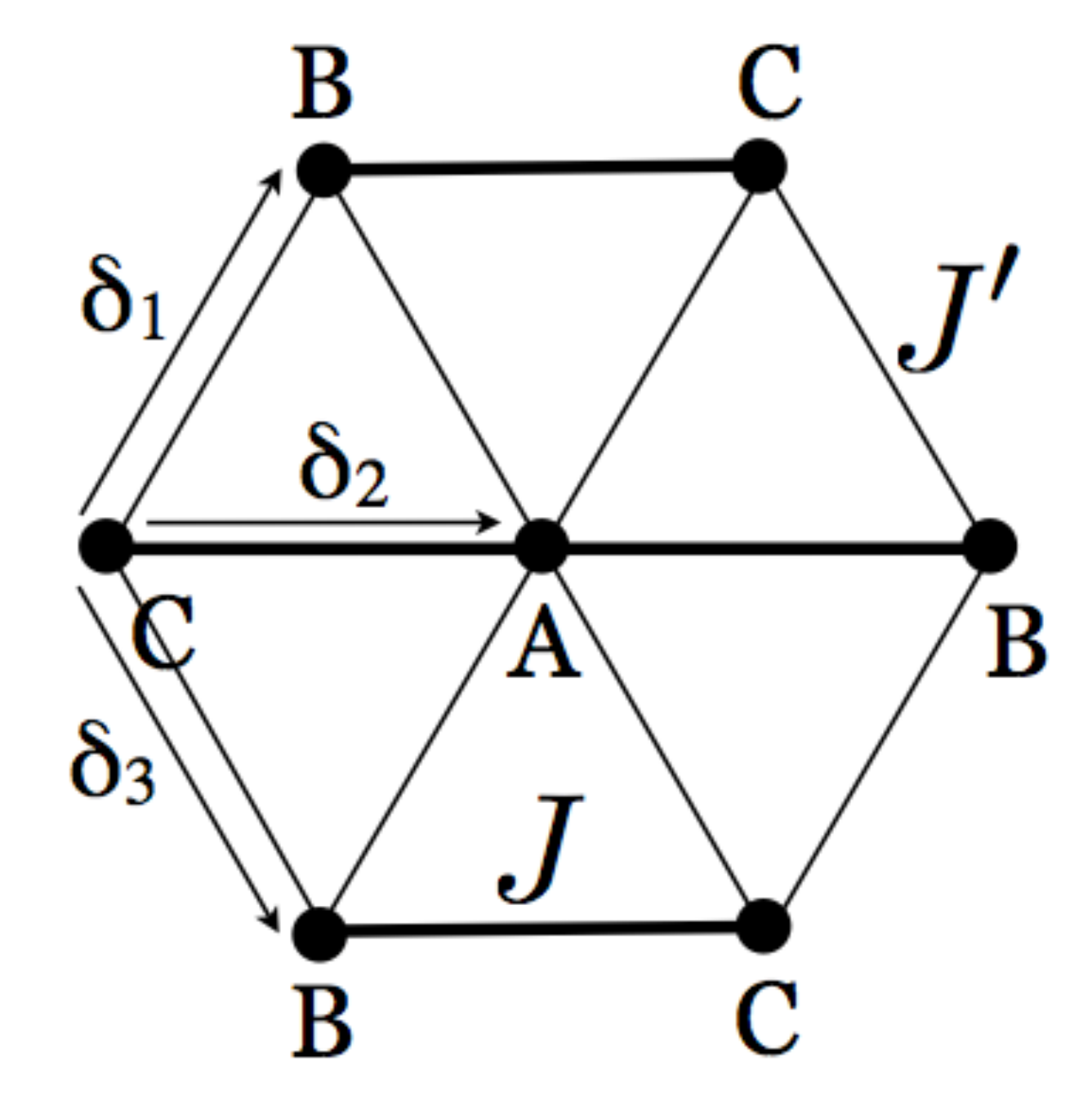}
\caption{Deformed triangular lattice. Solid (thin) lines denote bonds with exchange constant $J$ ($J'$) correspondingly. Also indicated are nearest-neighbor 
vectors $\delta_j$ as well as A, B and C sublattice structure.}
\label{fig:lattice}
\end{figure}

It turns out that a simple generalization of the triangular lattice Heisenberg model whereby exchange interactions on the nearest-neighbor
bonds of the triangular lattice take two different values -- $J$ on the horizontal bonds and $J'$ on the diagonal bonds, as shown in Figure~\ref{fig:lattice} -- 
leads to a very rich and not yet fully understood phase diagram which sensitively depends on the magnitude of the site spin $S$ and
magnetic field $h$. Such a distorted, or spatially anisotropic, triangular lattice model interpolates between simple unfrustrated square lattice ($J=0, J'\neq 0$),
strongly frustrated triangular lattice ($J=J'\neq 0$) and decoupled spin chains ($J\neq0, J'=0$). 

An unexpectedly large number of experimental systems seems to fit into this simple distorted triangular lattice model with spin $S=1/2$:
Cs$_2$CuCl$_4$ (shows extended spinon continuum, $J'/J = 0.34$), Cs$_2$CuBr$_4$ (shows magnetization plateau, $J'/J \approx 0.7 $) and 
Ba$_3$CoSb$_2$O$_9$ (shows magnetization plateau, $J'/J \approx 1$). Importantly, a number of very interesting organic Mott insulators of X[Pd(dmit)$_2$]$_2$
and $\kappa$-(ET)$_2$Z families can also be approximately described by the spatially anisotropic $J-J'$ model with additional ring-exchange (involving four spins 
and higher order) terms.
It is widely believed that these materials are weak Mott insulators in a sense of being close to a metal-insulator transition. As such, they are best
described by distorted $t - t' - U$ Hubbard model, so that spatial anisotropy of exchanges follows from that of single particle hopping parameters, $J'/J \sim (t'/t)^2$.

The (intentionally) rather narrow focus of the review leaves out several important recent developments.
Kagom\'e lattice antiferromagnets represent perhaps the most notable omission.
A lot is happening there, both in terms of experiments on materials such as herbertsmithite and volborthite \cite{Mendels2011} and
in terms of theoretical developments \cite{Yan2011,Han2012}. This very important area of frustrated magnetism deserves
its own review. We also have avoided another very significant area of development -- systems with significant spin-orbit interactions.
Progress in this area has been summarized in recent reviews \cite{Nussinov2013,Witczak-Krempa2013}.

The presentation is organized as follows: Section II contains brief review of the states of classical model in a magnetic field. Section III 
describes phase diagrams of the semi-classical $S\gg 1$ (Section III B) and $S=1/2$ (Section III C) models. 
Novel ordered states, a collinear SDW and a spin nematic, which are characterized by the absence of $S=1$ transverse spin excitations,
are described in Section IV. Section V summarizes key experimental findings relevant to the review, including
recent developments in organic Mott insulators. 
%Section VI lists several important, but not included in the review, topics and concludes the review.

\section{Classical model in a magnetic field}
\label{sec:classical}

Triangular antiferromagnets in an external magnetic field have been extensively studied for decades, and found to possess 
unusual magnetization physics that remains only partially understood. Underlying much of this interesting behavior is the discovery, 
made long ago,\cite{Kawamura1985} that in a magnetic field, Heisenberg spins with isotropic exchange interactions exhibit a large {\em accidental} classical ground-state 
degeneracy. That is, at finite magnetic fields, there exists an infinite number of continuously deformable classical spin configurations that 
constitute minimum energy states, but are {\em not} symmetry related. 

This degeneracy is understood by the observation that the Hamiltonian of the isotropic triangular lattice antiferromagnet in magnetic field ${\bf h}$ can
be written, up to an unessential constant, as

\begin{equation}
H_0 = \frac{J}{2} \sum_{\bf r} \Big[ {\bf S}_{\bf r} + {\bf S}_{\bf r + \delta_1} + {\bf S}_{\bf r + \delta_2} - \frac{{\bf h}}{3 J}\Big]^2 .
\label{eq:1}
\end{equation}

The sum is over all sites ${\bf r}$ of the lattice and nearest-neighbor vectors ${\bf \delta}_{1,2}$ are indicated in the Figure~\ref{fig:lattice}.
Note that Zeeman terms ${\bf S}_{\bf r} \cdot {\bf h}$ appear three times for every spin in this sum, which explains the factor of $1/3$ in the ${\bf h}$ term
in \eqref{eq:1}. We immediately observe that every spin configuration which nullifies every term in the sum \eqref{eq:1} belongs to the lowest 
energy manifold of the model. Given the {\em side-sharing} property of the triangular lattice, so that fixing all spins in one elementary triangle fixes  two spins
in each of the adjacent triangles, sharing sides with the first one, this implies that all such states exhibit a
three-sublattice structure and must satisfy
\begin{equation}
{\bf S}_A + {\bf S}_B + {\bf S}_C = \frac{{\bf h}}{3 J} .
\label{eq:2}
\end{equation}

This condition provides 3 equations for 6 angles needed to describe 3 classical unit vectors. In the absence of the field, the 3 undetermined angles can
be thought of as Euler's angles of the plane in which the spins spontaneously form a three-sublattice $120^\circ$ structure. However, this remarkable
feature persists for ${\bf h} \neq 0$ as well. There, the symmetry of the Hamiltonian \eqref{eq:1} is reduced to $U(1)$ but the degeneracy persists:
one of the free angles can be thought as gauge degree of freedom to rotate all spins about the axis of the field ${\bf h}$, while the remaining two
constitute the phenomenon of {\em accidental degeneracy}.

\begin{figure}[h]
\centering
\includegraphics[width=8.0cm]{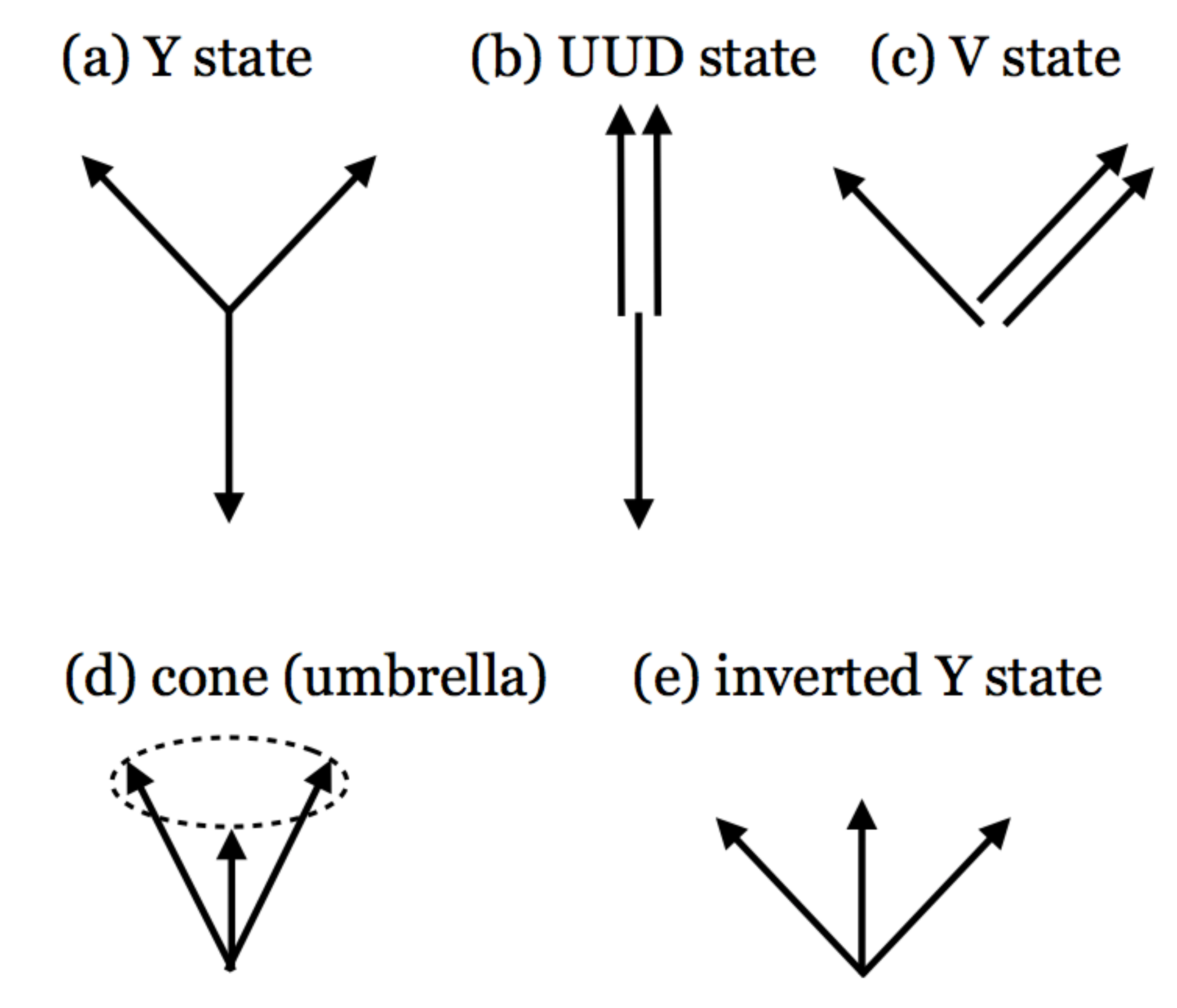}
\caption{Various spin configurations from the classical ground state manifold. (a) coplanar Y state, (b) collinear UUD (up-up-down), (c) coplanar V state,
(d) non-coplanar cone (umbrella) state, (e) inverted Y state. }
\label{fig:states}
\end{figure}

Remarkably, thermal (entropic) fluctuations lift this extensive degeneracy in favor of the two {\em coplanar} (Y and V states) and one {\em collinear} (UUD) spin
configurations, shown in Figure~\ref{fig:states}. Symmetry-wise, coplanar states break two different symmetries -- a discrete $Z_3$ symmetry, which corresponds to the
choice of sublattice on which the down spin (in the case of Y) or the minority spin (in the case of V) is located, and a continuous $U(1) = O(2)$ symmetry of rotations
about the field axis. The collinear UUD state breaks only the discrete $Z_3$ symmetry (a choice of sublattice for the down spin). 
Selection of these simple states out of infinitely many configurations, which satisfy \eqref{eq:2}, by thermal fluctuations
represents a textbook example of the `order-by-disorder' phenomenon \cite{Chalker2011}.

\begin{figure}[h]
\centering
\includegraphics[width=7.0cm]{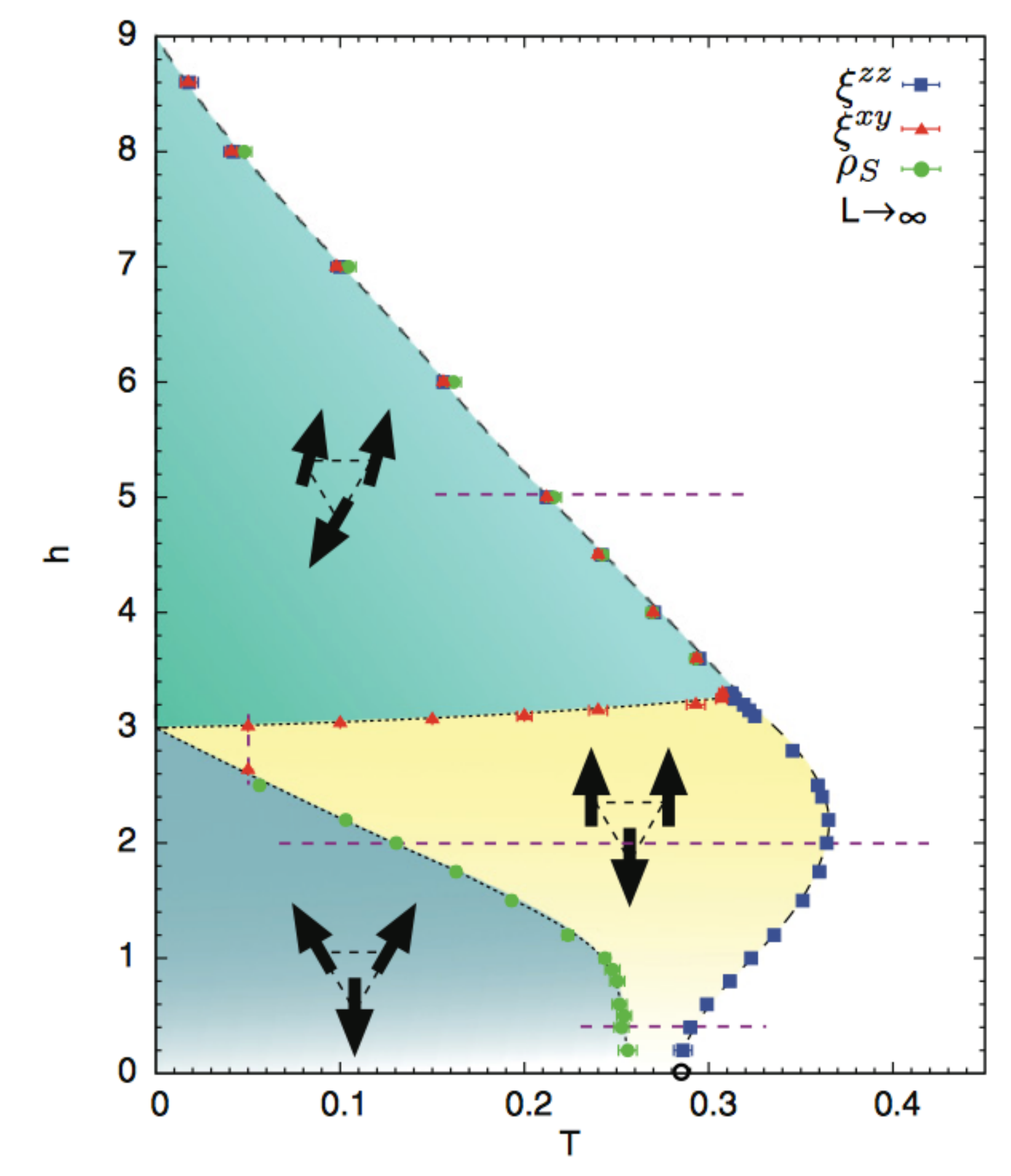}
\caption{Magnetic field phase diagram of the classical triangular lattice antiferromagnet. 
[Adapted from Seabra {\it et al.}, \prb {\bf  84}, 214418 (2011). Copyright  2011 by the American Physical Society.]
Transition points determined by the Monte Carlo simulations are shown by filled symbols. 
Continuous phase transitions are drawn with a dashed line, while Berezinskii-Kosterlitz-Thouless phase transitions are drawn with a dotted line. 
For fields $h \leq 3$ a double transition is found upon cooling from the paramagnet, while for $h \geq 3$ only a single transition is found.
Behavior of the phase transition lines in the low-field region $h\leq 0.2$, which is left unshaded in the diagram, is not settled at the present.
See Ref.\onlinecite{Kawamura2010} for the recent study of $h=0$ line.
%Adapted from Seabra {\it et al.}, \prb {\bf  84}, 214418 (2011). Copyright  2011 by the American Physical Society.
}
\label{fig:diagram}
\end{figure}

The resulting enigmatic phase diagram, first sketched by Kawamura and Miyashita in 1985,  Ref.~\onlinecite{Kawamura1985}, 
continues to attract much attention - and in fact remains
not fully understood. Figure \ref{fig:diagram} shows the result of recent simulations \cite{Seabra2011}, which
studied critical properties of various phase transitions in great detail, differs in important aspect from the original suggestion -- 
it is established now that there is no direct transition between the Y state and
the paramagnetic phase (very similar results were obtained in 
an extensive study \cite{Gvozdikova2011}). The two phases are separated by the intervening UUD state which extends down to lowest accessible field values
and before the transition to the paramagnetic state. Figure \ref{fig:diagram} also shows that of all entropically selected state, the UUD state
is most stable - it extends to higher $T$ than either $Y$ or $V$ states.

\begin{figure}[h]
\centering
\includegraphics[width=7.0cm]{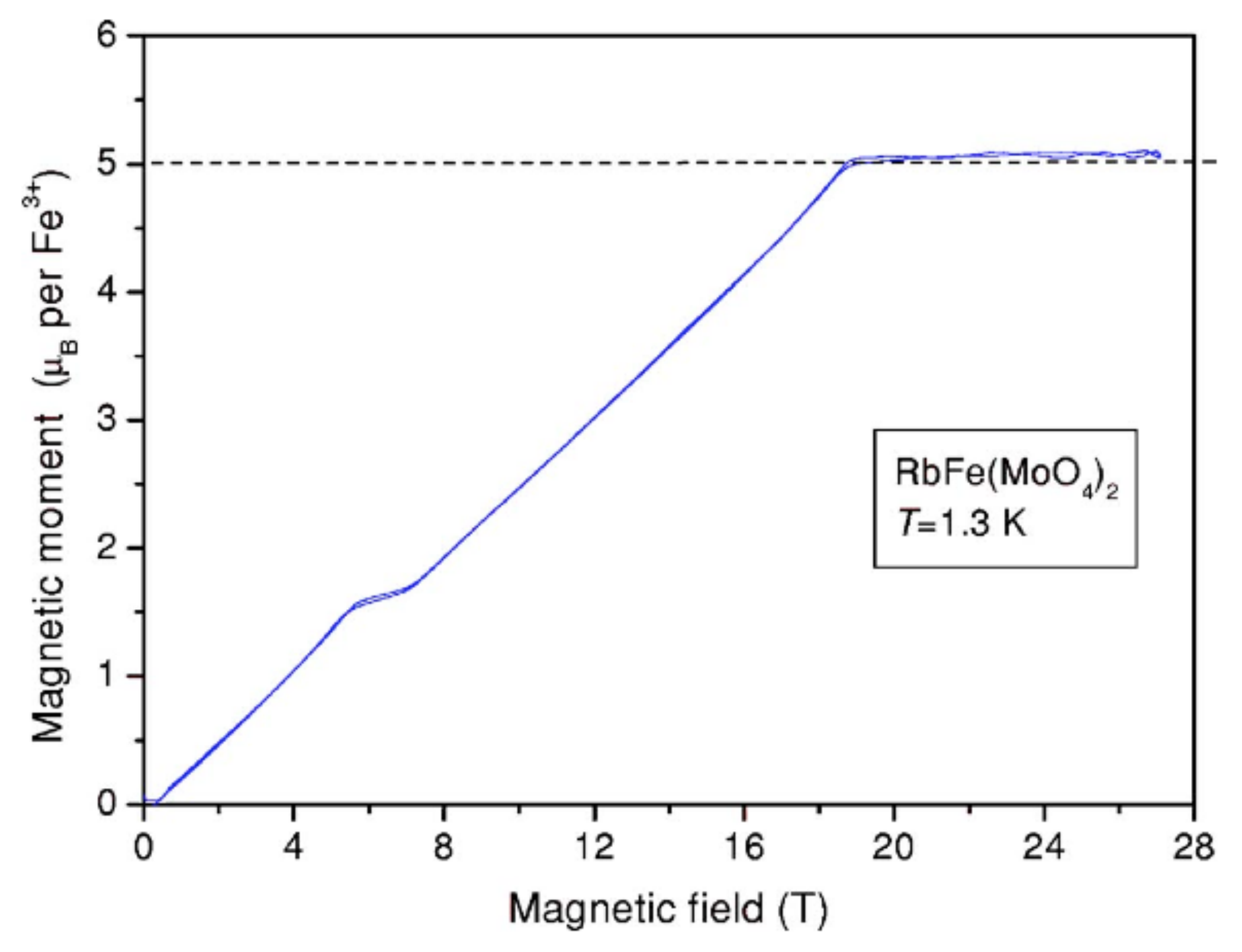}
\caption{Magnetization curve $M(h)$ of a spin-$5/2$ antiferromagnet RbFe(MoO$_4$)$_2$ at $T=1.3$K. 
[Adapted from Smirnov {\it et al.}, \prb {\bf  75}, 134412 (2007). Copyright  2007 by the American Physical Society.]}
\label{fig:Mh}
\end{figure}

The UUD is also the most `visible' of the three selected states - it shows up as a plateau-like feature in the magnetization curve $M(h)$,
see Figure~\ref{fig:Mh} which shows experimental data for a $S=5/2$ triangular lattice antiferromagnet RbFe(MoO$_4$)$_2$ \cite{Smirnov2007}. Notice that
a strict magnetization plateau at $1/3$ of the full (saturation) value $M_{\rm sat}$, $M = M_{\rm sat}/3$, is possible only in 
the quantum problem ({\it i.e.} the problem with finite spin $S$),
when all spin-changing excitations with $S^z \neq 0$ are characterized by {\em gapped} spectra, and at absolute zero $T=0$, when no
excitations are present in the ground state (see next Section~\ref{sec:QTLAF} for complete discussion).
At any finite $T$ thermally excited spin waves are present and lead to a finite, albeit different from the neighboring non-plateau states,
slope of the magnetization $M(h)$. In the case of the classical problem we review here the gap in spin excitation spectra is itself $T$-dependent
and disappears as $T\to 0$: as a result its magnetization `quasi-plateau' too disappears in the $T\to 0$ limit. However, at any finite $T$,
less than that of the transition to the paramagnetic state, the slope of the $M(h)$ for the UUD state is different from that of the Y and V states \cite{griset2011deformed}.

The most outstanding, and still not resolved issue, is the conjectured $SO(3)$ breaking transition \cite{Kawamura1984}
at $T_v \approx 0.285 J$ and $h=0$. The transition is driven by the proliferation, above the critical temperature $T_v$,
of the $Z_2$ vortices, which are defects of spin chirality \cite{Kawamura1984}. It is by now established that unlike the 
case of Berezinsky-Kosterlitz-Thouless vortex-unbinding transition \cite{Korshunov2006}, the spin correlation length remains finite (albeit very large)
below $T_v$, resulting in a disordered ``spin-gel" state \cite{Kawamura2010}. Whether the change from high-temperature vortex-dominated
regime to the low-temperature spin-fluctuation-dominated one is a true transition or a sharp crossover remains the topic of active 
debate \cite{Wintel1995,Southern1995,Azaria1992,Mouhanna2011,Hasselmann2013}. Experimental ramifications of this interesting scenario are
reviewed in \cite{Kawamura2011}.

A classical system with spatially anisotropic interactions offers an interesting generalization of the `order-by-disorder' phenomenon.
Consider slightly deformed triangular lattice, with $J' < J$ (see Fig.~\ref{fig:lattice}). An arbitrary weak deformation  lifts, at $T=0$, the accidental degeneracy in favor
of the simple non-coplanar umbrella state (configuration `d' in Fig.~\ref{fig:states}) in the whole range of $h$ below the saturation field. The energy gain of this well-known spin configuration is of the order $(J-J')^2/J$.
One thus can expect that, for sufficiently small difference $R = J-J'$, entropic fluctuations, which favor coplanar and collinear spin states at a finite $T$, can still 
overcome this classical energy gain and stabilize collinear and coplanar states {\em above} some critical temperature
which can be estimated as $T_{\rm umb-uud} \sim R^2/J$. As a result, the UUD phase  `decouples' from the $T=0$ axis.
The leftmost point of the UUD phase in the $h-T$ phase diagram now occurs at a finite $T_{\rm umb-uud}$ as was indeed observed in 
Monte Carlo simulation of the simple triangular lattice model in Ref.\onlinecite{griset2011deformed} (see Fig.10 of that reference), 
as well as in a more complicated ones, for models defined on deformed pyrochlore \cite{Pinettes2002} and 
Shastry-Sutherland \cite{Moliner2009} lattices.
It is worth noting that the roots of this behavior can be traced to the famous Pomeranchuk effect in $^3$He, where the crystal phase 
has higher entropy than the normal Fermi-liquid phase. As a result, upon heating, the liquid phase {\em freezes} into a solid \cite{Pomeranchuk1950,Richardson1997}.
In the present classical spin problem it is  
`superfluid' umbrella phase which freezes into a `solid' UUD phase upon heating 
(see discussion of the relevant terminology at the end of Section~\ref{sec:quant-iso} below).

\section{Quantum model in magnetic field}
\label{sec:QTLAF}
\subsection{Isotropic triangular lattice}
\label{sec:quant-iso}

Much of the intuition about selection of coplanar and collinear spin states by thermal fluctuations applies to the 
most interesting case of {\em quantum} spin model on a (deformed) triangular lattice. Only now it is quantum fluctuations (zero-point motion) 
which differentiate between different classically-degenerate (or, nearly degenerate) states and lift the accidental degeneracy. 

One of the first explicit calculations of this effect was carried out by Chubukov and Golosov \cite{chubukov1991quantum}, who used
semiclassical large-$S$ spin wave expansion in order to systematically separate classical and quantum effects.
This well-known technique relies on Holstein-Primakoff representation of spin operators in terms of bosons. The representation is nonlinear,
$S_{\bf r}^z = S - a_{\bf r}^\dagger a_{\bf r} , S_{\bf r}^+ = (2S - a_{\bf r}^\dagger a_{\bf r})^{1/2} a_{\bf r}$, and leads, 
upon expansion of square roots in powers of small parameter $1/S$, to bosonic spin-wave Hamiltonian 
$H = E_{\rm cl} + \sum_{k = 2}^\infty H^{(k)}$. Here $E_{\rm cl}$ stands for the classical energy of spin configuration, which scales as $S^2$, while
each of the subsequent terms $H^{(k)}$ are of $k$-th order in operators $a_{\bf r}$ and scale as $S^{2 - k/2}$.
Diagonalization of quadratic term $H^{(2)}$ provides one with the dispersion $\omega_{\bf k}^{(m)}$ 
of  spin wave excitations
(${\bf k}$ is the wave vector and $m$ is the band index), in terms of which quantum zero-point energy is given by
$\langle H^{(2)} \rangle = (1/2) \sum_{m, {\bf k}} \omega_{\bf k}^{(m)}$. This energy scales as $S$, and thus provides the leading
quantum correction to the classical ($\propto S^2$) result.

The main outcome of the calculation \cite{chubukov1991quantum} is the finding that quantum fluctuations too selects 
coplanar Y and V and collinear UUD states (states a, b and c in Figure~\ref{fig:states}) out of many other classically degenerate ones.
The authors also recognized the key special feature of the UUD state - being collinear, this state preserves $U(1)$ symmetry
of rotations about the magnetic field axis. The absence of broken continuous symmetry implies that spin excitations have a finite energy gap
in the dispersion. This expectation is fully confirmed by the explicit calculation \cite{chubukov1991quantum}  which finds 
that the gaps of the two low energy modes are given by $| h - h_{c1,c2}^0|$, where the lower/upper critical fields are 
given by $h_{c1}^0 = 3JS - 0.5JS/(2S)$ and 
$h_{c2}^0 = 3JS + 1.3JS/(2S)$, correspondingly. (The third mode describes a high-energy precession with energy $h S$.) 
The uniform magnetization $M$, being the integral of motion, remains at $1/3$ of the maximum (saturation) value, $M = M_{\rm sat}/3$,
in the UUD stability interval $h_{c1}^0 < h < h_{c2}^0$.

As a result, magnetization curve $M(h)$ of the quantum triangular lattice antiferromagnet is non-monotonic and exhibit striking 
1/3 {\em magnetization plateau} in the finite field interval $\Delta h = h_{c2}^0 - h_{c1}^0 = 1.8 JS/(2S)$. It is worth noting that 
this is not a narrow interval at all, $\Delta h/h_{\rm sat} = 0.2/(2S)$ in terms of the saturation field $h_{\rm sat} = 9 J S$.
Extending this large-$S$ result all way to the $S=1/2$ implies that the magnetization plateau takes up $20\%$ 
of the whole magnetic field interval $0 < h < h_{\rm sat}$.

Numerical studies of the plateau focus mostly on the quantum spin-1/2 problem (numerical studies of higher spins are much more `expensive')
and confirm the scenario outlined above.
The plateau at $M=M_{\rm sat}/3$ is indeed found, and moreover its width in magnetic field agrees very well
with the described above large-$S$ result, extrapolated to $S=1/2$. \cite{Honecker2004,Farnell2009,tay2010variational,sakai2011,chen2013,Hotta2013}

The pattern of symmetry breaking by the coplanar/collinear states is described by the following spin expectation values
\begin{eqnarray}
{\text{Y state,}} &&~ 0 < h < h_{c1}^0: \langle S_{\bf r}^+ \rangle = a e^{i \varphi} \sin[{\bf Q} \cdot {\bf r}] ,\nonumber\\ 
&&\langle S_{\bf r}^z \rangle = b - c \cos^2[{\bf Q} \cdot {\bf r}]; \label{eq:Y}\\
{\text{UUD state,}} &&~ h_{c1}^0 < h < h_{c2}^0: \langle S_{\bf r}^+ \rangle = 0, \nonumber\\
&& \langle S_{\bf r}^z \rangle = M - c \cos[{\bf Q} \cdot {\bf r}]; \label{eq:uud} \\
{\text{V state,}} &&~h_{c1}^0 < h < h_{\rm sat}: \langle S_{\bf r}^+ \rangle = a e^{i \varphi} \cos[{\bf Q} \cdot {\bf r}] , \nonumber\\
&& \langle S_{\bf r}^z \rangle = b - c \cos^2[{\bf Q} \cdot {\bf r}]. \label{eq:V}
\end{eqnarray}
Here the ordering wave vector ${\bf Q} = (4\pi/3, 0)$ is commensurate with the lattice which results in only three possible values
that the product ${\bf Q} \cdot {\bf r} = 2\pi \nu/3$ can take ($\nu = 0, 1, 2$), modulo $2\pi$. Angle $\varphi$ specifies orientation of the ordering plane for
coplanar spin configurations within the $x-y$ plane, $M$ is magnetization per site, and parameters $a,b,c$ are constants dependent upon the field magnitude.

{\bf Connection with `super' phases of bosons:}
Eq.\eqref{eq:uud} identifies UUD state as a collinear ordered state which can be thought of as {\em solid}. Its `density' $\langle S_{\bf r}^z \rangle$ is {\em modulated}
as $\cos[{\bf Q} \cdot {\bf r}]$, as appropriate for the solid, and as a result its local magnetization  follows
simple `up-up-down' pattern within each elementary triangle. It obviously respects $U(1)$ symmetry of rotations about $S^z$ axis.
The coplanar Y and V states break this $U(1)$ symmetry by spontaneously selecting angle $\varphi$. Note that in addition they 
are characterized by the modulated density $\langle S_{\bf r}^z \rangle$, which makes them {\em supersolids}: the superfluid order (magnetic order
in the $x-y$ plane as selected by $\varphi$) co-exists with the solid one (modulated $z$ component, or density).
This useful connection is easiest to make precise \cite{Murthy1997} in the case of $S=1/2$ when the following mapping between a {\em hard-core} lattice Bose
gas an a spin-1/2 quantum magnet can easily be established:
\begin{equation}
a^+_{\bf r} \leftrightarrow S^+_{\bf r}, a_{\bf r} \leftrightarrow S^-_{\bf r}, n_{\bf r} = a^+_{\bf r} a_{\bf r} \leftrightarrow S^{-}_{\bf r} - 1/2 .
\label{eq:super}
\end{equation}
The superfluid order is associated with finite $\langle a^+_{\bf r}\rangle$ while the solid one with modulated (with momentum ${\bf Q}$ in our notations)
boson density $\langle n_{\bf r}\rangle$.

\subsection{Spatially anisotropic triangular antiferromagnet with $J'\neq J$}
\label{sec:anisotropic-triangular}

Consider now a simple deformation of the triangular lattice which makes exchange interaction on diagonal bonds, $J'$, different
from those on horizontal ones $J$, so that $R = J - J' \neq 0$. 
This simple generalization of the Heisenberg model leads to surprisingly complicated and not yet fully
understood phase diagram in the magnetic field ($h$) - deformation ($R$) plane.

Semiclassical ($S \gg 1$) analysis of this problem is complicated by the fact that arbitrary small  $R\neq 0$ 
removes accidental degeneracy of the problem in favor of the unique non-coplanar and incommensurate cone (umbrella) 
state, already discussed in Sec.~\ref{sec:classical},  state `d' in Fig.~\ref{fig:states}.
This simple state gains energy of the order $\delta E_{\rm class} \sim S^2 R^2/J$ per spin. Its structure is described by 
\begin{equation}
\langle {\bf S}_{\bf r} \rangle = M \hat{z} + c ( \cos[{\bf Q}' \cdot {\bf r} + \varphi] \hat{x} + \sin[{\bf Q}' \cdot {\bf r} + \varphi] \hat{y}) ,
\label{eq:umbrella}
\end{equation}
where classically the ordering wave vector ${\bf Q}' = (2 \cos^{-1}[-J'/2J], 0)$ is a continuous function of $J'/J$.
Being non-coplanar, this state is characterized by the finite chirality 
$\chi \sim {\bf S}_{\bf r} \cdot {\bf S}_{\bf r + \delta_1} \times {\bf S}_{\bf r + \delta_2} \sim M(2\sin[{\bf Q}' \cdot \delta_1] -\sin[{\bf Q}' \cdot \delta_2]) $.

At the same time, for sufficiently small $R$, the quantum energy gain due to
zero-point motion of spins, which is of the order $\delta E_{\rm q} \sim S J$ per spin, should be able to overcome $\delta E_{\rm class}$
and still stabilize one of the coplanar/collinear states considered in Sec.~\ref{sec:quant-iso} above. Comparing the two contributions,
we conclude, following Ref.~\onlinecite{alicea2009quantum}, that the classical-quantum competition can be parameterized by the dimensionless
parameter $\delta \sim \delta E_{\rm class}/\delta E_{\rm q} \sim S (R/J)^2$. (In the following, we will use more precise value
$\delta = (40/3) S (R/J)^2$, with numerical factor $40/3$ as introduced in \cite{alicea2009quantum}
for technical convenience.)

Explicit consideration of this competition is rather difficult due to complicated dependence of the parameters of the coplanar states
on magnetic field $h$ and exchange deformation $R$. However, inside the $M=M_{\rm sat}/3$ magnetization plateau phase, the spin structure is actually pretty simple,
as equation \eqref{eq:uud} shows, which suggests that UUD state can be used as a convenient starting point for accessing more complicated states.
This approach, which amounts to the investigation of the {\em local stability} of the UUD phase, 
was carried out in \cite{alicea2009quantum}. Similar to \cite{chubukov1991quantum}, the calculation is based
on the three-spin UUD unit cell, resulting in three spin wave branches. One of these, describing total spin precession, is a high-energy mode
not essential for our analysis, while the two others, which describe relative fluctuations of spins, are the relevant low-energy modes.

\begin{figure}[h]
\centering
\includegraphics[width=8.5cm]{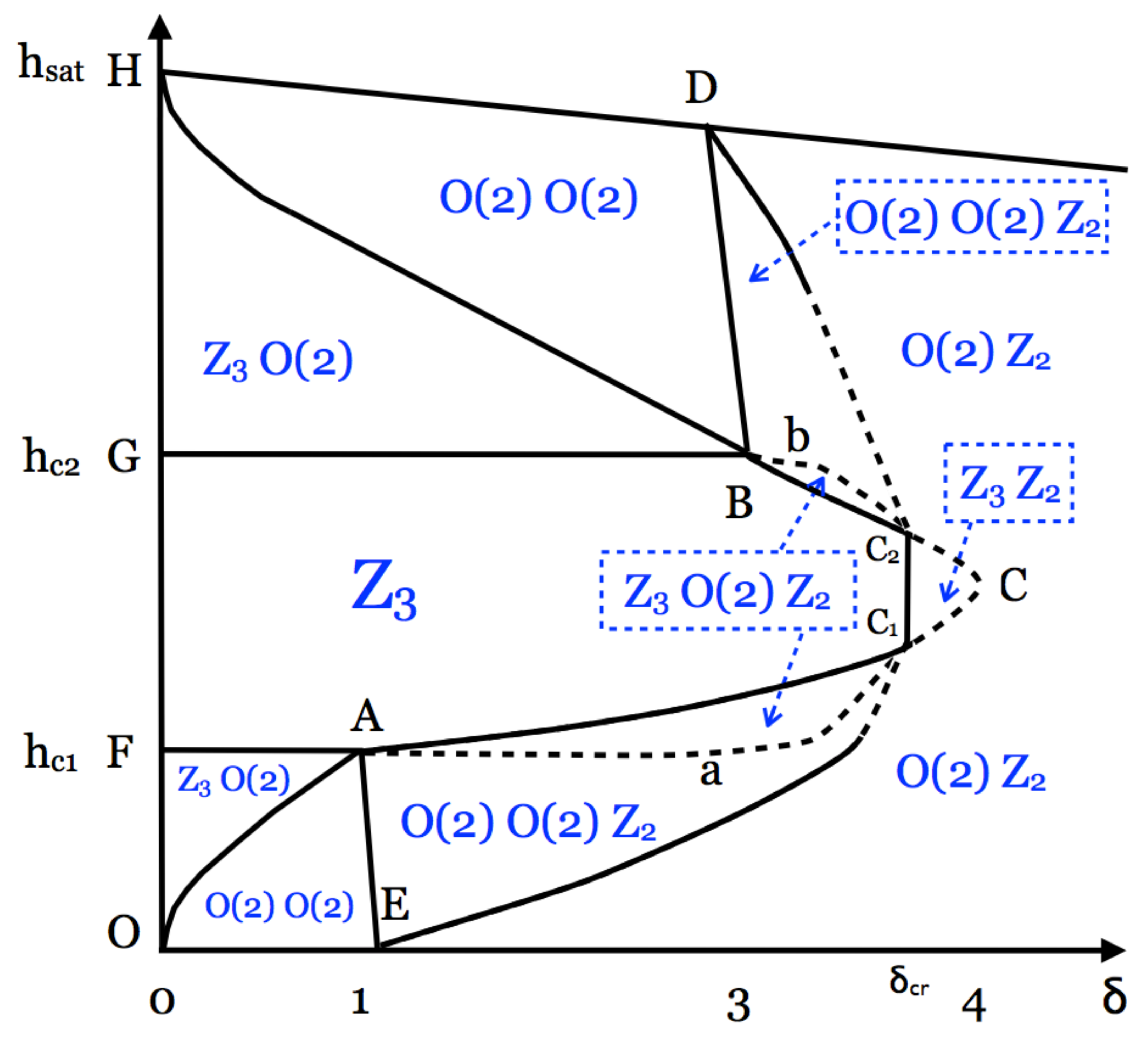}%{DiagramS3a}
\caption{Schematic large-$S$ phase diagram of the deformed triangular lattice antiferromagnet. Phase boundary lines are schematic and
serve to indicate possible shape only approximately. Dashed lines indicate location of conjectured phase transitions.
Capital letters A through O indicated various (multi)critical points discussed in the text.
Each phase is characterized by the set of broken symmetries, as indicated by (blue) symbols. Observe that the line H-B-b-C$_2$-C-C$_1$-a-A-O
is the transition line separating phases with broken $Z_3$ from those with broken O(2). This phase diagram is based on Refs. \onlinecite{alicea2009quantum,Chubukov2013,Starykh2014}.}
\label{fig:diag-S}
\end{figure}

Within the UUD plateau phase, which is bounded by FACBG lines in Figure~\ref{fig:diag-S}, both low energy spin wave modes remain gapped.
As discussed above, the only symmetry this collinear state breaks is $Z_3$.
The gaps of the low-energy spin wave modes (to be called mode 1 and mode 2 in the following) are
given by $|h - h_{c1,c2}|$, where now the lower and upper critical fields $h_{c1,c2}(\delta)$ depend on the dimensionless deformation
parameter $\delta$. Gap's closing at the lower $h_{c1}$ (upper $h_{c2}$) critical fields of the plateau implies Bose-Einstein condensation (BEC) of the appropriate 
magnon mode (1 or 2) and the appearance of the transverse to the field spin component $\langle S_{\bf r}^+\rangle \neq 0$, see \eqref{eq:Y} and \eqref{eq:V}.
This BEC transition breaks spin rotational $U(1)$ symmetry via spontaneous selection of the `superfluid' phase $\varphi$.
The spin structure of the resulting `condensed' state sensitively depends on the wave vector ${\bf k}_{1/2}$ of the condensed magnon.

Key results of the large-$S$ calculation in Refs.~\onlinecite{alicea2009quantum,Chubukov2013,Starykh2014} can now be summarized as follows:\\
({\bf a}) In the interval $0 < \delta \leq 1$ the lower critical field is actually independent of $\delta$, $h_{c1} = h_{c1}^0$, and the minimum
of the spin wave mode 1 remains at ${\bf k}_{1} = 0$. As a result, BEC condensation of mode 1 at $h = h_{c1}$ signals
the transition to the commensurate Y state, which lives in the region OAF in Figure~\ref{fig:diag-S}.
In addition to breaking the continuous $U(1)$ symmetry, the Y state inherits broken $Z_3$ from the UUD phase
(which corresponds to the selection of the sublattice for the {\em down} spins).\\
({\bf b}) For $1 < \delta \leq 4$, the low critical field increases and, simultaneously, the spin wave minima shift from zero 
to finite momenta $\pm {\bf k}_1 = (\pm k_1,0)$. Thus, at $h = h_{c1}$, the spectrum softens at two different wave vectors at the same time.
This opens an interesting possibility of the simultaneous coherent condensation of magnons with opposite momenta 
$+{\bf k}_1$ {\em and} $ -{\bf k}_1$.\cite{Nikuni1995}
It turns out, however, that repulsive interaction between condensate densities at $\pm {\bf k}_1$ makes this 
energetically unfavorable \cite{alicea2009quantum}. Instead, at the transition the symmetry between the two possible condensates
is broken spontaneously and magnons condense at a single momentum, $+{\bf k}_1$ {\em or} $ -{\bf k}_1$.
The resulting state, denoted as {\em distorted umbrella} in \cite{alicea2009quantum}, is characterized by the
broken $Z_3, U(1)$ {\em and} $Z_2$ symmetries -- the latter corresponds to the choice $+ {\bf k}_1$ or $- {\bf k}_1$.
This non-coplanar state lives in the narrow region bounded by lines AC$_1$ (solid) and AaC$_1$ (dashed) in Figure~\ref{fig:diag-S}.\\
({\bf c}) Similar developments occur near the upper critical field $h_{c2}$. In the interval $0 < \delta \leq 3$ the upper critical field is 
actually independent of $\delta$, $h_{c2} = h_{c2}^0$, and the minimum
of the spin wave mode 2 remains at ${\bf k}_{2} = 0$. The transition on the line GB is towards the coplanar and commensurate V state,
bounded by GBH in Figure~\ref{fig:diag-S}. This state is characterized by broken $Z_3 \times U(1)$ symmetries.\\
({\bf d}) In the interval $3 < \delta \leq 4$, the upper field $h_{c2}$ diminishes and simultaneously the minimum of the mode 2 
shifts from zero momentum to the two degenerate locations at $\pm {\bf k}_2 = (\pm k_2,0)$. Here, too, at the condensation transition
(along the line BC$_2$) it is energetically preferable to break $Z_2$ symmetry between the two condensates and to spontaneously
select just one momentum, $+{\bf k}_2$ {\em or} $ -{\bf k}_2$. This leads to {\em distorted umbrella}  
with broken $Z_3 \times U(1) \times Z_2$ symmetries. This state lives between (solid) BC$_2$ and (dashed) BbC$_2$ lines in Figure~\ref{fig:diag-S}. \\
({\bf e}) {\em Spin nematic region.} The critical fields $h_{c1}$ and $h_{c2}$ merge at the plateau's end-point $\delta = 4$ (point C). 
The minima of the spin wave
modes 1 and 2 coincide at this point ${\bf k}_1 = {\bf k}_2 = (k_0, 0)$ with $k_0 = \sqrt{3/(10S)}$. 
The end-point of the UUD phase thus emerges as a point of an extended symmetry hosting 
four linearly-dispersing gapless spin wave modes \cite{alicea2009quantum} (two branches, each gapless at $(\pm k_0, 0)$). 
Single-particle analysis of possible instabilities at the plateau's 
end-point (point C, $\delta =4$) shows that in addition to the expected $U(1)\times U(1)$ symmetry (the two $U(1)$'s represent phases of the single particle
condensates), the state at $\delta =4$ point posses an unusual $P_1$ symmetry -- the magnitude of the condensate at this point is
not constrained \cite{alicea2009quantum}. The enhanced degeneracy of the plateau's end point C can also be 
understood from the observation that the two chiral distorted umbrella states
merging at the point C are characterized, quite generally, by the {\em different} chiralities: Condensation of magnons at $h_{c1,c2}$,
described in items ({\bf b}) and ({\bf c}) above, proceed independently of each other - hence the chiralities of the two states
are not related in any way. Thus the merging point of the two phases, point C, must possess enhanced symmetry.

\begin{figure}[h]
\centering
\includegraphics[width=8.5cm]{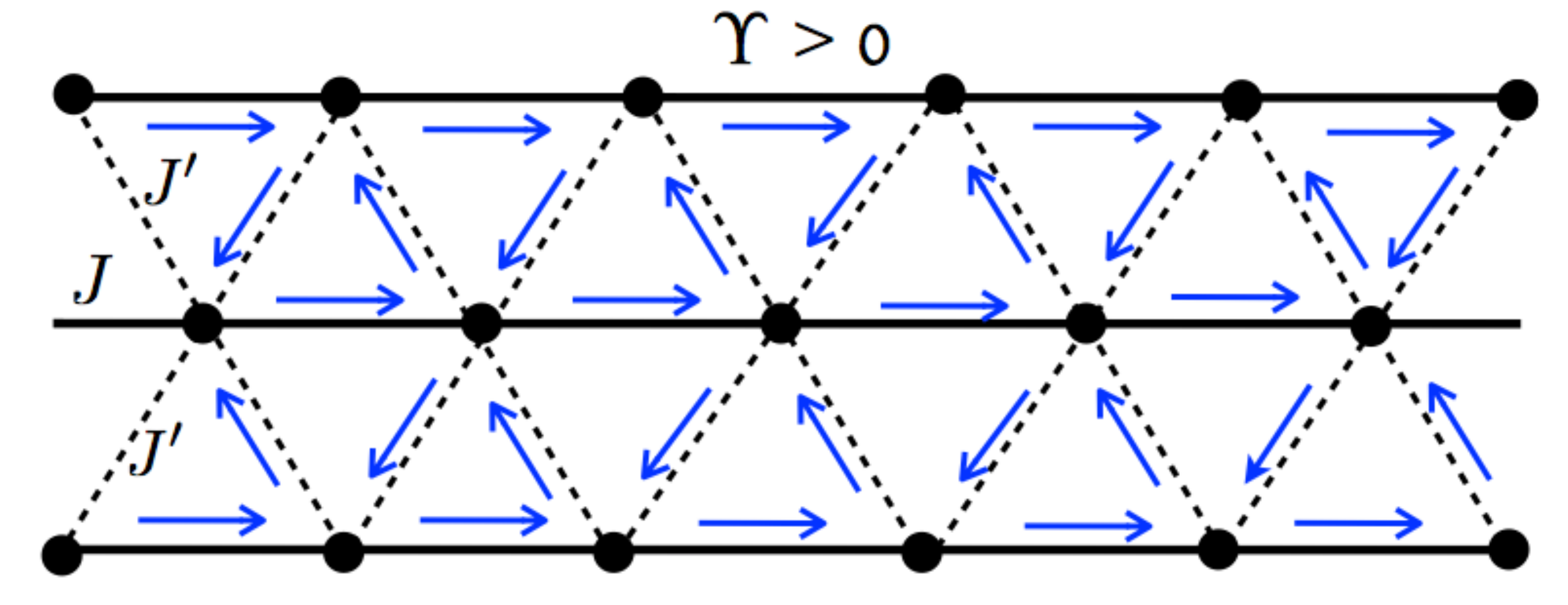}%{DiagramS2a}
\caption{Pattern of spin currents in the spin nematic phase, for $\Upsilon > 0$.}
\label{fig:spincurrent}
\end{figure}

Unconstrained magnitude mode hints at a possibility of a {\em two-particle condensation} - and indeed recent work \cite{Chubukov2013} 
has found that a single-particle instability at $\delta=4$ is pre-empted by the two-particle one at a slightly smaller value of 
$\delta = \delta_{\rm cr} = 4 - O(1/S^2)$.
This is indicated by the (dashed) line C$_1$-C$_2$  in the Figure. This happens via the development of the `superconducting'-like instability
of the magnon pair fields $\Psi_{1, {\bf p}} = d_{1,+{\bf k}_0 + {\bf p}} d_{2,-{\bf k}_0 - {\bf p}}$ and
$\Psi_{2, {\bf p}} = d_{1,-{\bf k}_0 + {\bf p}} d_{2,+{\bf k}_0 - {\bf p}}$ and consists in the appearance of two-particle condensates
$\langle \Psi_{1, {\bf p}}\rangle = \langle \Psi_{2, {\bf p}}\rangle = i \Upsilon/|{\bf p}|$. The sign of the real-valued Ising order parameter 
$\Upsilon$ determines the sense of spin-current circulation on the links of the triangular lattice, as illustrated in Figure~\ref{fig:spincurrent}.
The spin current is defined as the 
ground state expectation value of the vector product of neighboring spins. For example, spin current on the AC link 
of the elementary triangle is given by
${\cal J}^z_{AC} = \hat{z} \cdot \langle {\bf S}_A \times {\bf S}_C \rangle \sim \Upsilon$.
In addition to spin currents, this novel state also supports finite spin chirality, 
$\langle {\bf S}_A \cdot {\bf S}_B \times {\bf S} \rangle \sim \Upsilon$, even though $\langle S^{x,y}_{A/B/C} \rangle = 0$ for each of the spins
individually.
At the same time, in 
the absence of single-particle condensation, $\langle d_{1/2}\rangle =0$,
the usual two-point spin correlation function $\langle S^a({\bf r}) S^b(0)\rangle$ is not affected by the two-particle 
$\langle \Psi_{1/2}\rangle \neq 0$ condensate: its transverse ($a=b = x$ or $y$) components continue to decay exponentially because of the
finite energy gap in the single magnon spectra, while the longitudinal ($a=b = z$) components continue to show a perfect UUD crystal order.

Hence the resulting state, which lives inside triangle-shaped region C$_1$-C-C$_2$ in Figure~\ref{fig:diag-S}, is a {\em spin-nematic} state.
It can also be called a {\em spin-current} state \cite{Chubukov2013}.
It is uniquely characterized by the sign of $\Upsilon$ which determines the sense (clockwise or counterclockwise) of spin current circulation.
The spin-current nematic is an Ising-like phase with massive excitations, which are domain walls separating domains of oppositely circling spin currents.
It is characterized by broken $Z_3 \times Z_2$, where $Z_2$ is the sign of the two-magnon order parameter $\Upsilon$ in the ground state. 
It is useful to note that spontaneous 
selection of the circulation direction can also be viewed as a spontaneous breaking of the spatial inversion symmetry ${\cal I}: x\to -x$, 
which changes direction of spin currents on all bonds. (A different kind of nematic is discussed in Section~\ref{sec:nematic}.)

\noindent
({\bf f}) {\em high-field region}, $h_{c2}(\delta) \ll h \leq h_{\rm sat}(\delta)$.
The high field region, $h \approx  h_{\rm sat}(\delta)$, can be conveniently analyzed within powerful Bose-Einstein condensation (BEC)
framework. This follows from the simple observation that the ground state of the spin-$S$
quantum model becomes a simple fully polarized state once the magnetic field is greater than the saturation field, $h > h_{\rm sat}(\delta)$
(which itself is a function of spin $S$ and exchange deformation $R$). Excitations above this exact ground state are standard spin waves
minimal energy for creating which is given by $h - h_{\rm sat}$. Similar to the situation near plateau's critical field $h_{c1/c2}$, these spin
waves are characterized by non-trivial dispersion with two degenerate minima at momenta $\pm {\bf Q}'$ (see expression below 
\eqref{eq:umbrella}).

Spontaneous condensation in one of the minima, which constitutes breaking of $Z_2$ symmetry, 
results in the usual {\em cone}, or umbrella, state which is characterized
by the spin pattern \eqref{eq:umbrella} which breaks spin-rotational $O(2)$. 
This state is realized to the right of the D-C$_2$ line in Figure~\ref{fig:diag-S}.

Instead, simultaneous  condensation of the high-field magnons in {\em both} minima results in the 
{\em coplanar} state \cite{Nikuni1995} (also known as fan state) of the $V$ type.
For any $\delta\neq 0$ the wave vector ${\bf Q}'$ is not commensurate with the lattice, which makes the coplanar state
to be {\em incommensurate} as well.
The condensates $\Psi_{1,2}= \sqrt{\rho} e^{i\theta_{1,2}}$ at wave vectors $\pm {\bf Q}'$ have equal magnitude $\sqrt{\rho}$
and each breaks $U(1)$ symmetry. By choosing the phases $\theta_{1,2}$, the resulting state breaks {\em two} $O(2)$ symmetries.
In total, the coplanar state is characterized by the broken $O(2) \times O(2)$.
It is instructive to think of these symmetries in a slightly different way - as of those associated with the total $\theta_+ = \theta_1 + \theta_2$ and
the relative $\theta_- = \theta_1 - \theta_2$ phases. The spin structure of this state is described by the incommensurate version of \eqref{eq:V}
which can be obtained by identifying $\varphi = \theta_+$ and replacing $\cos[{\bf Q}\cdot {\bf r}] \to \cos[{\bf Q}'\cdot {\bf r} + \theta_{-}]$.
The latter replacement shows that the relative $O(2)$ symmetry, associated with the phase $\theta_{-}$, can also 
be thought of as a {\em translational} symmetry associated with the shift of the spin configuration by the vector ${\bf r}_0$ such that
${\bf Q}'\cdot {\bf r}_0 = \theta_{-}$, modulo $2\pi$.

Moving to the left, we come to the special {\em commensurate} point, $\delta=0$ at $h=h_{\rm sat}$. Here ${\bf Q}' \to {\bf Q} = (4\pi/3, 0)$, 
resulting in the commensurate coplanar V state. As noted right below Eq.~\eqref{eq:V}, here ${\bf Q}\cdot {\bf r} = 2\pi \nu/3$, 
with $\nu =0,1,2$, making the V state a three-sublattice one. Continuous $O(2)$ translational symmetry of the incommensurate V phase
is replaced here by the discrete $Z_3$ symmetry. In other words, the relative phase $ \theta_{-}$ is now restricted to the set of three equivalent
(up to a global translation of the lattice) degenerate values. The $Z_3 \to O(2)$ transition across the line H-B between the two coplanar states  
is thus of a commensurate-incommensurate transition (CI) type. It is described by the classical two-dimensional sine-Gordon model
with the nonlinear $\cos[3\theta_{-}]$ potential describing the locking of the relative phase to the $Z_3$ set. \cite{chen2013}
In the vicinity of point H in the diagram the line of the CI transition follows $h_{\rm sat} - h \sim \sqrt{\delta}$. \cite{Starykh2014}
While the arguments presented here are valid in the immediate vicinity of $h_{\rm sat}$, the identification of the whole line H-B
as the CIT line between the commensurate and incommensurate V states is possible due to the additional evidence reported in
item ({\bf c}) above -- commensurate V state is reached from the UUD state in the whole interval $0 < \delta \leq 3$.

The next task is to connect the incommensurate coplanar V state, which occupies region HDB in Figure~\ref{fig:diag-S}, with 
the incommensurate cone state, to the right of D-C$2$ line. Since the V state has equal densities of bosons in the $\pm {\bf Q}'$
points, while the cone has finite density only in one of them, continuous transition between these two states at finite condensate density
(that is, at any $h < h_{\rm sat}$) is {\em not possible}. At infinitesimally small condensate density, {\it i.e.} at $h = h_{\rm sat}$, direct transition
is possible - it occurs at point D, which is a point of extended $O(2)\times O(2)\times O(2)$ symmetry: the two $O(2)$'s are phase symmetries
while the third one is an emergent symmetry associated with the invariance of the potential energy at the constant total condensate density, 
$\rho = \rho_1 + \rho_2$,
with respect to the distribution of condensate densities $\rho_{j=1,2} = \Psi^\dagger_j \Psi_j$ between the $\pm {\bf Q}'$ momenta. 
Large-$S$ calculation of ladder diagrams which describe quantum corrections to the condensate energy place point D
at $\delta=2.91$ \cite{Starykh2014}.
Assuming that the first order transition
does not realize, we are forced to conclude that V and cone states must be separated by the intermediate phase, occupying DBbC$_2$ region.
This phase breaks the $O(2)$ symmetry between $\rho_1$ and $\rho_2$ and interpolates smoothly between the symmetric situation 
$\rho_1=\rho_2$, on the D-B line, and the asymmetric one $\rho = \rho_1 $ and $\rho_2 =0$ (or vice versa) on the line D-C$_2$.
In doing so the momentum of the `minority' condensate is found to evolve continuously from the initial ${\bf q}_2 = {\bf Q}'$ (which coincides with
the momentum of the `majority' condensate $\rho_1$) on the line D-C$_2$
to the final ${\bf q}_2 = - {\bf Q}'$ on the D-B line \cite{Starykh2014}. The resulting phase is a non-coplanar one, with strongly pronounced
asymmetry in the $x-y$ plane: 
$\langle S^+_{\bf r}\rangle = \sqrt{\rho_1} e^{i\theta_1} e^{i {\bf Q}'\cdot{\bf r}} + \sqrt{\rho_2} e^{i\theta_2} e^{i {\bf q}_2\cdot{\bf r}}$.
The state is characterized by the broken $O(2)\times O(2)\times Z_2$.
For the lack of better term we call it {\em double spiral} \cite{Starykh2014}.

Going down along the field axis takes us toward the dashed B-b-C$_2$ line below which, according to the analysis summarized in item (c) above,
represents a phase with broken $Z_3 \times Z_2 \times O(2)$. Hence along this line $Z_3$ is replaced by $O(2)$, which makes it a continuation
of the C-IC transitions line H-B.

\noindent
({\bf g}) {\em low-field region}, $0 \leq h \ll h_{c1}(\delta)$. Semiclassical analysis at zero field $h=0$ is well established and predicts incommensurate
spiral state with zero total magnetization $M=0$ of course. Quantum fluctuations renormalize strongly parameters 
of the spin spiral \cite{Weihong1999}. The most quantum case of the spin $S=1/2$ remains not fully understood even for the relatively weak deformation
of exchanges $R=J - J' \leq J$ and is described in more details in Section~\ref{sec:spin1/2}.

At $\delta=0$ one again has commensurate three-sublattice antiferromagnetic state, widely known as a $120^\circ$ structure, which evolves into 
commensurate Y state in external magnetic field $h \neq 0$. Phenomenological analysis of Ref.\onlinecite{chen2013}, Section III E, shows Y state becomes 
incommensurate when the deformation $R$ exceeds $R_c \sim h^{3/2}$. Analysis near $h_{c1}$, reported in (b), tells that C-IC transition line must end up
at point A. Comparing the energies of the incommensurate coplanar Y and the incommensurate umbrella state in the limit of vanishing 
magnetic field $h\to 0$, described in \cite{alicea2009quantum}, identifies point E at $\delta=1.1$ and $h=0$ as the point of the transition
between the $O(2)\times O(2)$ [the incommensurate coplanar V] and the $O(2)\times Z_2$ [the incommensurate cone] breaking states.
Thus point E is analogous to point D. 

By the arguments similar to those in part ({\bf f}) above, there must be an intermediate phase with broken $O(2)\times O(2)\times Z_2$.
It occupies region E-C$_1$-a-A in Figure~\ref{fig:diag-S}. The state between A-C$_1$ and A-a-C$_1$ lines is characterized
by different broken symmetries (see item (b) above) which makes the line A-a-C$_1$ to be the line of the $Z_3 \to O(2)$ transition.
It thus has to be viewed as a continuation of the CIT line O-A.

To summarize, the quasi classical phase diagram in Figure~\ref{fig:diag-S} contains many different phases. It worth keeping in mind
that it has been obtained under assumption of {\em continuous} phase transitions between states with different orders. Several of the
shown there phase boundaries are tentative -- their existence is conjectured based on different symmetry properties of the states 
they are supposed to separate. To highlight their conjectured nature, such lines are drawn {\em dashed} in Figure~\ref{fig:diag-S}.
These include line B-b-C$_2$ which separates distorted umbrella (with broken $Z_3\times O(2) \times Z_2$) and double spiral (with broken 
$O(2) \times O(2) \times Z_2$), and similar to it line A-a-C$_1$ located right below the UUD phase. The end-points of these dashed
lines are conjectured to be C$_2$ and C$_1$, correspondingly, which are the points of the two-magnon condensation [item ({\bf e})].
Line D-C$_2$, separating phases with broken $O(2) \times O(2) \times Z_2$ and $O(2) \times Z_2$, established via high-field analysis
in item ({\bf f}), is conjectured to end at the same C$_2$. Since different behavior cannot be ruled out at the present, its extension
to the near-plateau region is indicated by the dashed line as well. Similar arguments apply to the line E-C$_1$. Finally, line C$_2$-C-C$_1$, covering
the very tip of the UUD plateau phase, is made dashed because it is located past the two-magnon condensation transition (line C$_1$-C$_2$) 
into the spin-nematic state instabilities of which have not been explored in sufficient details yet.

One of the most unexpected and remarkable conclusions emerging from the analysis summarized here is the identification of the continuous 
line of C-IC transitions (line H-B-b-C$_2$-C-C$_1$-a-A-O), separating phases with discrete $Z_3$ from those with continuous $O(2)$ symmetry.
Its existence owes to the non-trivial interplay between geometric frustration and quantum spin fluctuations in the triangular antiferromagnet.

{\bf Spin excitation spectra:} Many of the ordered {\em non-collinear} states described above harbor spin excitations with rather unusual characteristics.
It has been pointed out some time ago \cite{Starykh2006,Chernyshev2006} that local non-collinearity of the magnetic order
results in the strong renormalization of the spin wave spectra at $1/S$ order. (This should be contrasted with the case of the collinear magnetic order,
where quantum corrections to the excitation spectrum appear only at $1/S^2$ order.)
This interesting effect, reviewed in \cite{Zhitomirsky2013}, is responsible for dramatic flattening of spin wave
dispersion and/or appearance of `roton-like' minima and related thermodynamic anomalies at temperatures as low as $0.1 - 0.2 J$ \cite{Zheng2006,Mezio2012}.

\subsection{Spin 1/2 spatially anisotropic triangular antiferromagnet with $J'\neq J$}
\label{sec:spin1/2}

We now turn to the case of most quantum system: a spin $1/2$  antiferromagnet. Qualitatively, one expects quantum fluctuations to be most pronounced
in this case, which suggests, in line with `order-by-disorder' arguments of Section~\ref{sec:quant-iso}, a selection of the ordered Y, UUD and V states
at and near the isotropic limit $J' \approx J$. Behavior away from this isotropic line represents a much more difficult problem, mainly due to 
the absence of physically motivated small parameter, which would allow for controlled analytical calculations. Aside from the two limits where small parameters
do appear, namely the high field region near the saturation field and the limit of weakly coupled spin chains (see below), the only available approach
is numerical.

Most of recent numerical studies of triangular lattice antiferromagnets focus on the zero field limit, $h=0$, and on the phase diagram as
a function of the ratio $J'/J$. These studies agree that a two-dimensional magnetic spiral order, well established at the isotropic $J'=J$ point, 
becomes incommensurate with the lattice when $J' \neq J$ and persists
down to approximately $J' = 0.5 J$. The ordering wave vector of the spiral ${\bf Q}'$ is strongly renormalized by quantum fluctuations \cite{Weihong1999,SRWhite2011}
away from the semiclassical result.
Below about $J' = 0.5 J$, strong finite size effects  and limited numerical accuracy of the exact 
diagonalization \cite{heidarian2009spin} and DMRG \cite{weng2006spin,SRWhite2011} methods does not allow one to obtain a definite answer about 
the ground state of the spin-1/2 $J-J'$ Heisenberg model. 

\begin{figure}[h]
\centering
\includegraphics[width=8.8cm]{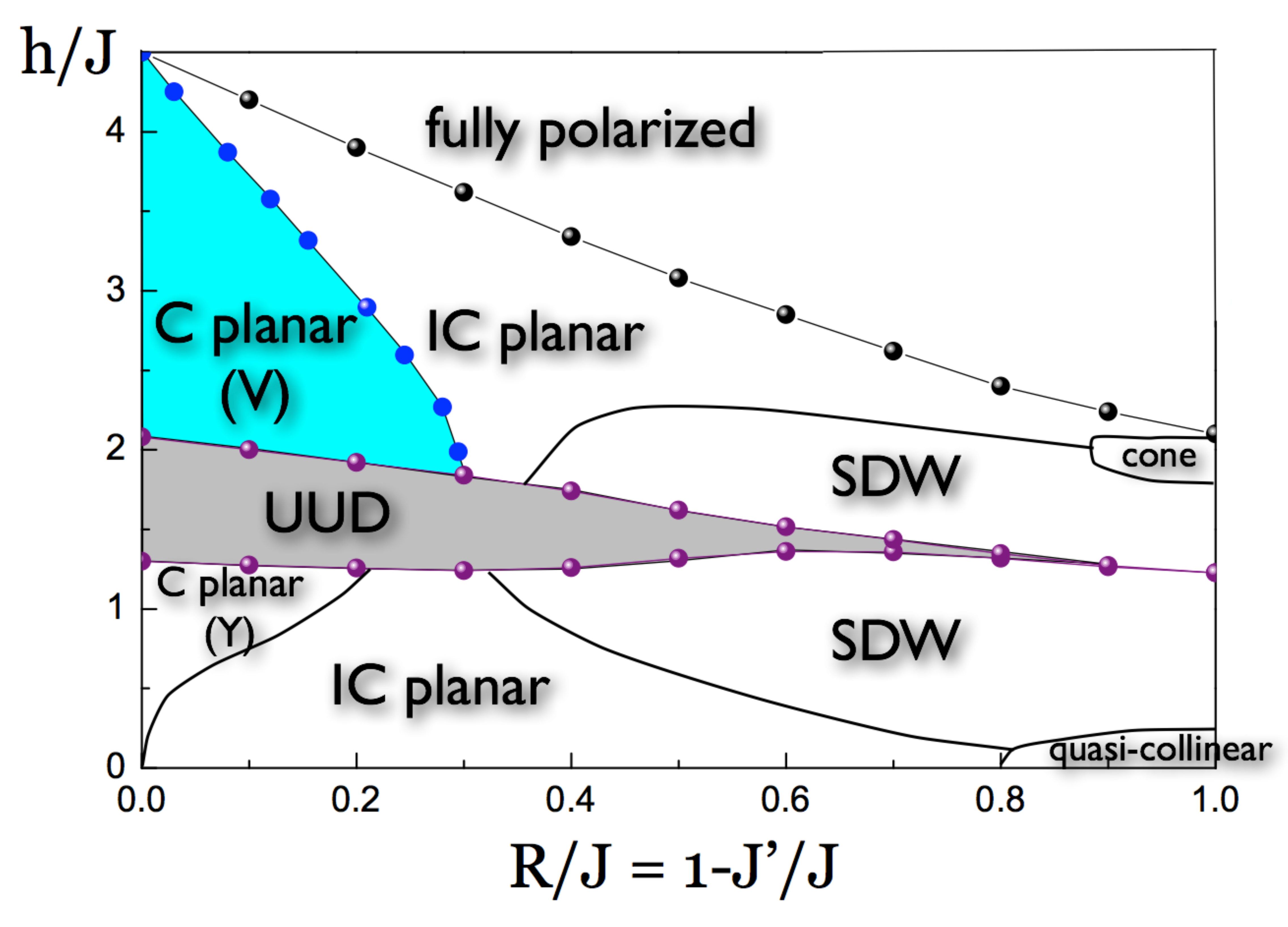}%{DiagramS2a}
\caption{Phase diagram of the spin $S=1/2$ spatially anisotropic triangular antiferromagnet in magnetic field. Vertical axis - magnetic field $h/J$, 
horizontal - dimensionless degree of spatial anisotropy, $R/J = 1-J'/J$. Notation C/IC stands for commensurate/incommensurate phases, correspondingly. 
Adapted from Chen {\it et al.}, Phys. Rev. B 87, 165123 (2013).}
\label{fig:diag1/2}
\end{figure}

This unexpected behavior is a direct consequence of the strong frustration inherent in the triangular geometry. In the $J' \ll J$ limit the lattice decouples into
a collection of linear spin chains weakly coupled by the frustrated interchain exchange 
${\cal H}' = J'\sum_{x, y} {\bf S}_{x,y} \cdot ({\bf S}_{x-1/2,y+1} + {\bf S}_{x+1/2,y+1})$. 
Even classically, spin-spin correlations between spins
from different chains are strongly suppressed as can be seen from the limit $J'/J \to 0$ when classical spiral wave vector 
$Q'_x = 2 \cos^{-1}[-J'/2J] \to \pi + J'/J$. In this limit the relative 
angle between the spin at (integer-numbered) site $x$ of the $y$-th chain and its neighbor at (half-integer-numbered) $x+1/2$ site of the $y+1$-th chain
approaches $\pi/2 + J'/(2J)$. Thus the scalar product of two classical spins at neighboring chains vanishes as $J'/(2J) \to 0$.

Quantum spins adds strong quantum fluctuations to this behavior, resulting in the numerically observed near-exponential decay of the inter-chain
spin correlations, even for intermediate value $J'/J \lesssim 0.5$  \cite{weng2006spin,SRWhite2011}. While some of the studies 
interpret such effective decoupling as the evidence of the spin-liquid ground state \cite{weng2006spin,heidarian2009spin}, the others conclude that the
coplanar spiral ground state persists all the way to $J'=0$ \cite{Pardini2008,SRWhite2011}.

Analytical renormalization group approach \cite{starykh2007ordering}, which utilizes symmetries and algebraic correlations of low-energy 
degrees of freedom of individual spin-1/2 chains, finds that the system experiences quantum phase transition from the ordered spiral state
to the unexpected collinear antiferromagnetic (CoAF) ground state. This novel magnetically ordered ground state is
stabilized by strong quantum fluctuations of critical spin chains. This finding is supported by the coupled-cluster study \cite{Bishop2009,Bishop2010},
functional renormalization group \cite{Reuther2011} as well as by the combined DMRG and analytical RG studies in \cite{Ghamari2011}.
It is fair to say that more studies of the very difficult $J'/J \to 0$ limit of the spatially anisotropic triangular model are highly desirable
in order to definitively sort out the issue of the ultimate ground state. 

Having described the limiting behavior along $h=0$ and $J'=J$ ($R=0$) axes, we now discuss the full $h-R$ phase diagram of the spin-1/2
Heisenberg model shown in Figure~\ref{fig:diag1/2} ($R = J - J'$). The diagram is derived from extensive DMRG study of triangular cylinders (spin tubes) 
composed of 
3, 6 and 9 chains and of lengths 120 - 180 sites (depending on $R$ and the magnetization value) as well as detailed
analytical RG arguments applicable in the  limit $J' \ll J$ \cite{chen2013}. 
It compares well, in the regions of small and intermediate $R$, 
with 
%the only other comprehensive study of the full anisotropy-magnetic field phase diagram - 
the variational and exact diagonalization
study by Tay and Motrunich \cite{tay2010variational}. The comparison is less conclusive in the regime of large anisotropy, $R \to 1$, which is
most challenging for numerical techniques as already discussed above. (The biggest uncertainty of the diagram in Figure~\ref{fig:diag1/2} consists in so far undetermined
region of stability of the cone phase and, to a lesser degree, the phase boundaries between incommensurate (IC) planar and SDW phases.)

The main features of the phase diagram of the quantum spin-1/2 model are:

(1) High-field incommensurate coplanar (incommensurate V or fan) phase, which is characterized by the broken $O(2)\times O(2)$ symmetry, 
is stable for {\em all} values of the exchange anisotropy $1 \geq R \geq 0$. This novel analytical finding, confirmed in DMRG simulations,
is described in Ref.\onlinecite{chen2013}. This result is specific to the quantum $S=1/2$ model -- for any other value of the site spin $S \geq 1$ there 
is a quantum phase transition between the incommensurate planar and the incommensurate cone phases at some $R_S$. The 
 critical value $R_S$ is spin-dependent and decreases monotonically with $S$. We find $R_S \approx 0.9, 0.5, 0.4$ for $S=1, 3/2, 2$,
respectively \cite{chen2013}.

(2) 1/3 Magnetization plateau (UUD phase) is present for all values of $R$ too: it extends from $R=0$ all the way to $R=1$. This striking conclusion
is based on analytical calculations near the isotropic point \cite{alicea2009quantum}, discussed in the previous Section, complementary 
field-theoretical calculations near the decoupled chains limit of $R \approx 1$ \cite{starykh2010extreme,chen2013} and on extensive 
DMRG studies of the UUD plateau in Ref.\onlinecite{chen2013}.

(3) A large portion of the diagram in Figure~\ref{fig:diag1/2}, roughly to the right of $R=0.5$, is occupied by the novel incommensurate collinear SDW phase.
Physical properties of this magnetically ordered and yet intrinsically quantum state are summarized in Section~\ref{sec:sdw} below. 

Comparing the quantum phase diagram in Figure~\ref{fig:diag1/2} with the previously described large-$S$ phase diagram in 
Figure~\ref{fig:diag-S}, one notices that both the high-field incommensurate coplanar 
and the UUD phases are present there too. The fact that both of these states become {\em more stable} in the spin-1/2 case and expand to 
the whole range of exchange anisotropy $0 < R < 1$, represents
a striking {\em quantitative} difference between the large-$S$ and $S=\frac{1}{2}$ cases. It should also be noticed that both phase diagrams 
demonstrate that the range of stability of the incommensurate cone (umbrella) phase
is greatly diminished.

In contrast, a collinear SDW phase, which occupies a good portion of the quantum phase diagram in Figure~\ref{fig:diag1/2},
is not present in Figure~\ref{fig:diag-S}  at all - and this constitutes a major qualitative distinction between the large-$S$ and the quantum $S=\frac{1}{2}$ cases.

\section{SDW and nematic phases of spin-1/2 models}

\subsection{SDW}
\label{sec:sdw}

The {\em collinear SDW phase} is characterized by the {\em modulated} expectation value of the local magnetization
\begin{equation}
\langle S^z_{\bf r}\rangle = M + {\text{Re}}[{\bm \Phi} e^{i {\bf k}_{\rm sdw} \cdot {\bf r}}] , 
\label{eq:sdw}
\end{equation} 
where ${\bm \Phi}$ is the SDW order parameter, and SDW wave vector ${\bf k}_{\rm sdw}$ is generally incommensurate
with the lattice and, moreover, is the function of the uniform magnetization $M$ and anisotropy $R$.
Eq.\eqref{eq:sdw} is very unusual for a classical (or semi-classical) spin system, where magnetic moments tend to behave as vectors of fixed length.
It is, however, a relatively common phenomenon in itinerant electron systems with nested Fermi surfaces \cite{Gruner1988,Gruner1994,Monceau2012}. 
The appearance of such a state in a {\em frustrated} system of coupled spin-1/2 chains is rooted in a deep
similarity between Heisenberg spin chain and one-dimensional spin-1/2 Dirac fermions \cite{Affleck1988,gogolin2004bosonization,starykh2005anisotropic}: 
thanks to the well-known 
phenomenon of one-dimensional spin-charge separation, the spin sectors of these two models are identical. Ultimately, it is this correspondence 
that is responsible for the `softness' of the amplitude-like fluctuations underlying the collinear SDW state of Figure~\ref{fig:diag1/2}.

\begin{figure}[h]
\centering
\includegraphics[width=7.cm]{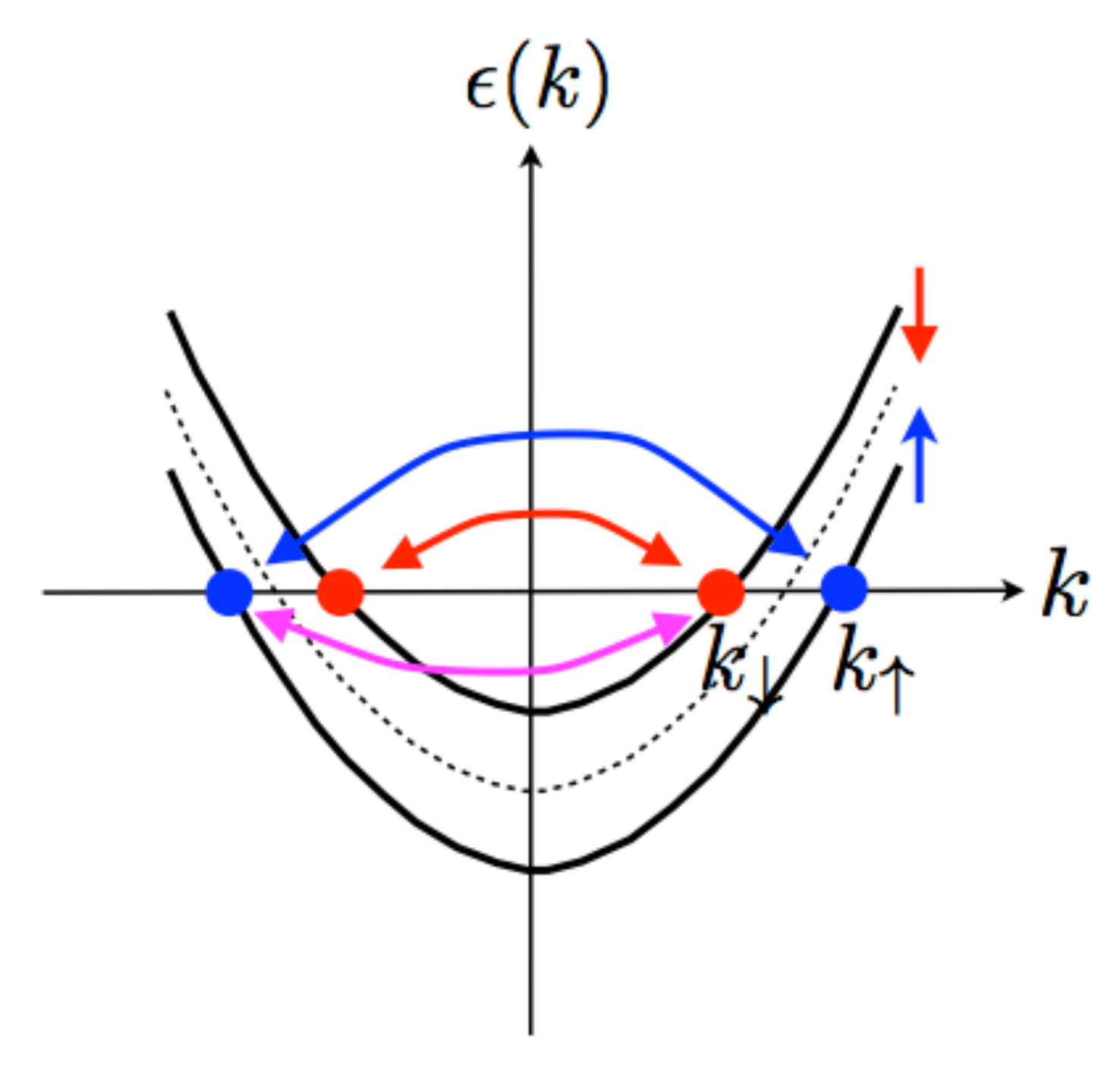}%{DiagramS2a}
\caption{Schematics of spinon dispersion for one-dimensional spin chain in magnetic field. Dashed line shows dispersion for zero field.
$k_{\uparrow} (k_\downarrow)$ denote Fermi momentum of spin 'up' ('down') spinons correspondingly. Fermi momentum in the absence 
of the field is $k_F = \pi/2$.}
\label{fig:sdw}
\end{figure}

Figure~\ref{fig:sdw} schematically shows a dispersion of $S=1/2$ electron in a magnetic field. Fermi-momenta of up- and down-spin electrons 
$k_{\uparrow/\downarrow}$ are shifted from the Fermi momentum of non magnetized chain, $k_F = \pi/2$, by $\pm \Delta k_F = \pm \pi M$,
where $M$ is the magnetization. As a result, the momentum of the {\em spin-flip} scattering processes, 
which determine transverse spin correlation function $\langle S^+ S^-\rangle$, is given by $\pm (k_\uparrow - ( - k_\downarrow)) = \pm 2 k_F = \pm \pi$, and
remains commensurate with the lattice. At the same time, longitudinal spin excitations, which preserve $S^z$, now involve momenta 
$\pm 2k_{\uparrow/\downarrow} = \pi (1 \pm 2M)$ and become incommensurate with the lattice. 
To put it differently, in the magnetized chain with $M\neq 0$ low-energy longitudinal spin fluctuations 
can be parameterized as $S^z(x) \sim {\cal S}^z e^{i (\pi - 2\pi M) x} + {\cal S}^{z *} e^{i (- \pi + 2\pi M) x}$, where (calligraphic) ${\cal S}^z(x)$
represents slow (low-energy) longitudinal mode. Similarly, transverse spin fluctuations are written as $S^+(x) \sim {\cal S}^+ e^{i \pi x}$,
with ${\cal S}^+$ representing low-energy transverse mode.

This simple fact has dramatic consequences for frustrated interchain interaction 
${\cal H}' = J'\sum_{x, y} {\bf S}_{x,y} \cdot ({\bf S}_{x-1/2,y+1} + {\bf S}_{x+1/2,y+1})$. 
Longitudinal ($z$) component of the {\em sum} of two neighboring spins on chain $(y+1)$ adds up to 
$S^z_{x-1/2,y+1} + S^z_{x+1/2,y+1} \to \sin[\pi M]({\cal S}^z_{x,y+1} e^{i (\pi - 2\pi M) x} + \text{h.c.})$, while the sum of 
their transverse components becomes a {\em derivative} of the smooth component of the transverse field ${\cal S}^+$, 
$S^+_{x-1/2,y+1} + S^+_{x+1/2,y+1} \to e^{i \pi x} \partial_x {\cal S}^+_{x,y+1}$. Hence the low-energy 
limit of the inter-chain interaction reduces to 
${\cal H}' \to \sum_y \int dx \{J' \sin[\pi M] {\cal S}^{z *}_{x,y} {\cal S}^z_{x,y+1} + J' {\cal S}^-_{x,y} \partial_x {\cal S}^+_{x,y+1} +\text{h.c.}\}$.
The presence of the spatial derivative in the second term severely weakens it \cite{starykh2010extreme} and results in
the domination of the density-density interaction (first term) over the transverse one (second term).
The field-induced shift of the Fermi-momenta from its commensurate value, $k_F \to k_{\uparrow/\downarrow}$,
together with frustrated geometry of inter-chain exchanges,
are the key reasons for the field-induced stabilization of the two-dimensional longitudinal SDW state.

Symmetry-wise, the SDW state breaks no global symmetries (time reversal symmetry is broken by the magnetic field, which also selects the $z$ axis), 
and, in particular, it preserves $U(1)$ symmetry of rotations about the field axis. This crucial feature implies the absence of  the off-diagonal magnetic
order $\langle S^{x,y}_{\bf r}\rangle = 0$ and would be gapless (Goldstone) spin waves.
Instead, SDW breaks lattice translational symmetry. Its order parameter ${\bm \Phi} \sim \langle {\cal S}^z\rangle \neq 0$ is determined by
inter-chain interactions \cite{starykh2010extreme,Starykh2013}.
Provided that ${\bf k}_{\rm sdw} = \pi (1 - 2M)\hat{x}$ is {\em incommensurate} with the lattice, the
%Consequently, its 
only low energy mode is expected to be the pseudo-Goldstone acoustic mode of broken translations, 
known as a {\em phason}. (Inter-chain interactions do affect ${\bf k}_{\rm sdw}$ but in the limit of small $J'/J$ this can be neglected.)
The phason is a purely longitudinal mode corresponding to the phase of the complex order parameter 
${\bm \Phi}$ and hence represents a modulation of $S^z$ only.  
This too is unusual in the context of insulating magnets, where, typically, the low energy collective modes are {\em transverse} spin waves, 
associated with small rotations of the spins away from their ordered axes. In the spin wave theory, longitudinal modes are typically 
expected to be highly damped \cite{Affleck1992,Schulz1996,Zhitomirsky2013}, and hence hard to observe. 
(For a recent notable exception to this rule see a study of amplitude-modulated magnetic state of PrNi$_2$Si$_2$ \cite{Blanco2013}.)
In the SDW state, the longitudinal phason mode in the only low energy excitation. Transverse spin excitations, which SDW also supports, 
have a finite spectral gap. This, in fact, is one of key experimentally identifiable features of the SDW phase.
More detailed description 
of spin excitations of this novel phase, as well as of the spin-nematic state reviewed below, can be found in the recent study \cite{Starykh2013}.

At present, there are three known routes to the field-induced longitudinal SDW phase for a quasi-one-dimensional system of weakly coupled
spin chains. The first, reviewed above, relies on the geometry-driven frustration of the transverse inter-chain exchange, which disrupts 
usual transverse spin ordering and promotes incommensurate order of longitudinal $S^z$ components. The other route, 
described in the subsection~\ref{sec:sdw-ising} below, relies on Ising anisotropy of individual chains.
Lastly, it turns out that a two-dimensional SDW state may also emerge in a system of weakly coupled {\em nematic} spin chains - this unexpected
possibility is reviewed in the subsection \ref{sec:nematic}.

\subsubsection{SDW in a system of Ising-like coupled chains}
\label{sec:sdw-ising}

There is yet another surprisingly simple route to the two-dimensional SDW phase. It consists in replacing Heisenberg chains with XXZ ones with
pronounced Ising anisotropy. It turns out that a sufficiently strong magnetic field, applied along the $z$ (easy) axis, drives individual chains
into a critical Luttinger liquid state with dominant {\em longitudinal}, $S^z-S^z$, correlations \cite{Okunishi2007}. This crucial property ensures that
weak residual inter-chain interaction selects incommensurate longitudinal SDW state as the ground state of the anisotropic two-dimensional system.

We note, for completeness, that not every field-induced gapless spin state is characterized by
the dominant longitudinal spin correlations. For example, another well-known gapped system, spin-1 Haldane chain, 
can too be driven into a critical Luttinger phase by sufficiently strong magnetic field \cite{Affleck1990,Sachdev1994,Zheludev2002,Essler2004}. 
However, that critical phase is instead dominated
by strong transverse spin correlations \cite{konik02,fath03}. As a result, a 2d ground state of weakly coupled spin-1 chains is a
usual cone state \cite{mcculloch08}.

\subsubsection{Magnetization plateau as a commensurate collinear SDW phase}

The {\em commensurate} case, when the wavelength $\lambda_{\rm sdw} = 2\pi/ k_{\rm sdw}$ is a rational
fraction of the lattice period, $\lambda_{\rm sdw} = q/p$, requires special consideration. (Here the lattice period is set to be $1$ and 
$q$ and $p$ are integer numbers.) Such commensurate state is possible at commensurate magnetization values $M^{(p,q)} = \frac{1}{2} (1 - \frac{2 q}{p})$.
At these values, `sliding' SDW state locks-in with the lattice, resulting in the loss of continuous translational symmetry.
The SDW-plateau transition is then an incommensurate-commensurate transition of the sine-Gordon variety \cite{starykh2010extreme,chen2013}.

It turns out that in two-dimensional triangular lattice such locking is possible, provided that integers $p$ and $q$ have the same
parity (both even or both odd) \cite{starykh2010extreme} (and, of course, provided that the spin system {\em is} in the two-dimensional collinear SDW phase). 
This condition selects $M=1/3 ~M_{\rm sat}$ plateau ($q=1, p=3$) as the most stable one,
in a sense of the biggest energy gap with respect to creation of spin-flip excitation (which changes total magnetization of the system by $\pm 1$).
The next possible plateau is at $M=3/5 ~M_{\rm sat}$ ($q=1, p=5$) \cite{starykh2010extreme} -- however this one apparently does not realize in the phase diagram
in Figure~\ref{fig:diag1/2}, perhaps because it is too narrow and/or happen to lie inside the (yet not determined numerically) cone phase.

Applied to the one-dimensional spin chain, the above condition can be re-written as a particular $S=\frac{1}{2}$ version of the Oshikawa-Yamanaka-Affleck condition
\cite{Oshikawa1997,Oshikawa2003} for the period-$p$ magnetization plateau in a spin-$S$ chain, $p S (1 - M/M_{\rm sat}) = {\text{integer}}$.
Interestingly, this shows that $p=3$ plateau at $M=1/3 ~M_{\rm sat}$ of the total magnetization $M_{\rm sat}$ is possible for all values of the spin $S$:
the quantization condition becomes simply $2 S = {\text{integer}}$. This rather non-obvious feature has in fact been numerically confirmed in several 
extensive studies \cite{okunishi03,Heidrich-Meisner2007,hikihara2010}.

\subsection{Spin nematic}
\label{sec:nematic}

{\em Spin nematic} represents another long-sought type of exotic ordering, Out of many possible nematic states \cite{Andreev1984,Chandra1991},
our focus here is on bond-nematic order associated with the two-magnon pairing \cite{Chubukov1991} and the 
appearance of the non-local order parameter $Q_{--} = S_{\bf r}^{-} S_{{\bf r}'}^{-}$
defined on the $\langle {\bf r},{\bf r}'\rangle$ bond connecting sites ${\bf r}$ and ${\bf r}'$. Such  order parameter can be build from quadrupolar 
operators $Q_{x^2-y^2} = S_{\bf r}^{x} S_{{\bf r}'}^{x} - S_{\bf r}^{y} S_{{\bf r}'}^{y}$ and $Q_{xy} = S_{\bf r}^{x} S_{{\bf r}'}^{y} + S_{\bf r}^{y} S_{{\bf r}'}^{x}$,
as $Q_{--} = Q_{x^2-y^2} - i Q_{xy}$ \cite{Hikihara2008}. This bond-nematic order is possible in both $S \geq 1$, where quadrupolar order
was originally suggested \cite{Blume1969}, and $S=1/2$ systems of localized spins, coupled by exchange interaction.

The magnon pairing viewpoint, explored in great length in \cite{Chandra1990,Hikihara2008,mzh2010}, is extremely useful for understanding basic properties of the
spin-nematic state: the nematic can be thought of as a `bosonic superconductor' formed as a result of two-magnon condensation 
$\langle S_{\bf r}^{-} S_{{\bf r}'}^{-}\rangle = \langle Q_{--} \rangle \neq 0$. As in a superconductor, a two-magnon condensate breaks $U(1)$ 
symmetry, which in this case is a breaking of the spin rotational symmetry with respect to magnetic field direction. 
It does not, however, break time-reversal symmetry
(which requires a single-particle condensation).
Just as in a superconductor,
single-particle excitations of the nematic phase are gapped. 
This implies that transverse spin correlation function $\langle S_{\bf r}^{+} S_{0}^{-}\rangle \sim e^{-r/\xi}$, which probes single magnon excitations,
is short-ranged and decays exponentially. At the same time, fluctuations of magnon density, which are probed by 
longitudinal spin correlation function $\langle S_{\bf r}^{z} S_{0}^{z}\rangle$, are sound-like acoustic (Bogoliubov) modes.

\subsubsection{Weakly coupled nematic chains}
\label{sec:nematic-chains}

Basic ingredients of this picture - gapped magnon excitations and attractive interaction between them - are nicely realized in the
spin-1/2 quasi-one-dimensional material LiCuVO$_4$, reviewed in Section~\ref{sec:sdw-exp}: 
the gap in the magnon spectrum is caused by the strong external magnetic field $h$ 
which exceeds (single-particle) condensation field $h_{\rm sat}^{(1)}$, while the attraction between magnons is caused by the {\em ferromagnetic} (negative)
sign of exchange interaction $J_1$ between the nearest spins of the chain. Under these conditions, the two-magnon bound state,
which lies below the gapped single particle states, condenses at $h_{\rm sat}^{(2)}$, which is higher than $h_{\rm sat}^{(1)}$.
As a result, a spin nematic state is naturally realized in each individual chain 
in the intermediate field interval $h_{\rm sat}^{(1)} < h < h_{\rm sat}^{(2)}$. \cite{mzh2010}

\begin{figure}[h]
\centering
\includegraphics[width=8.cm]{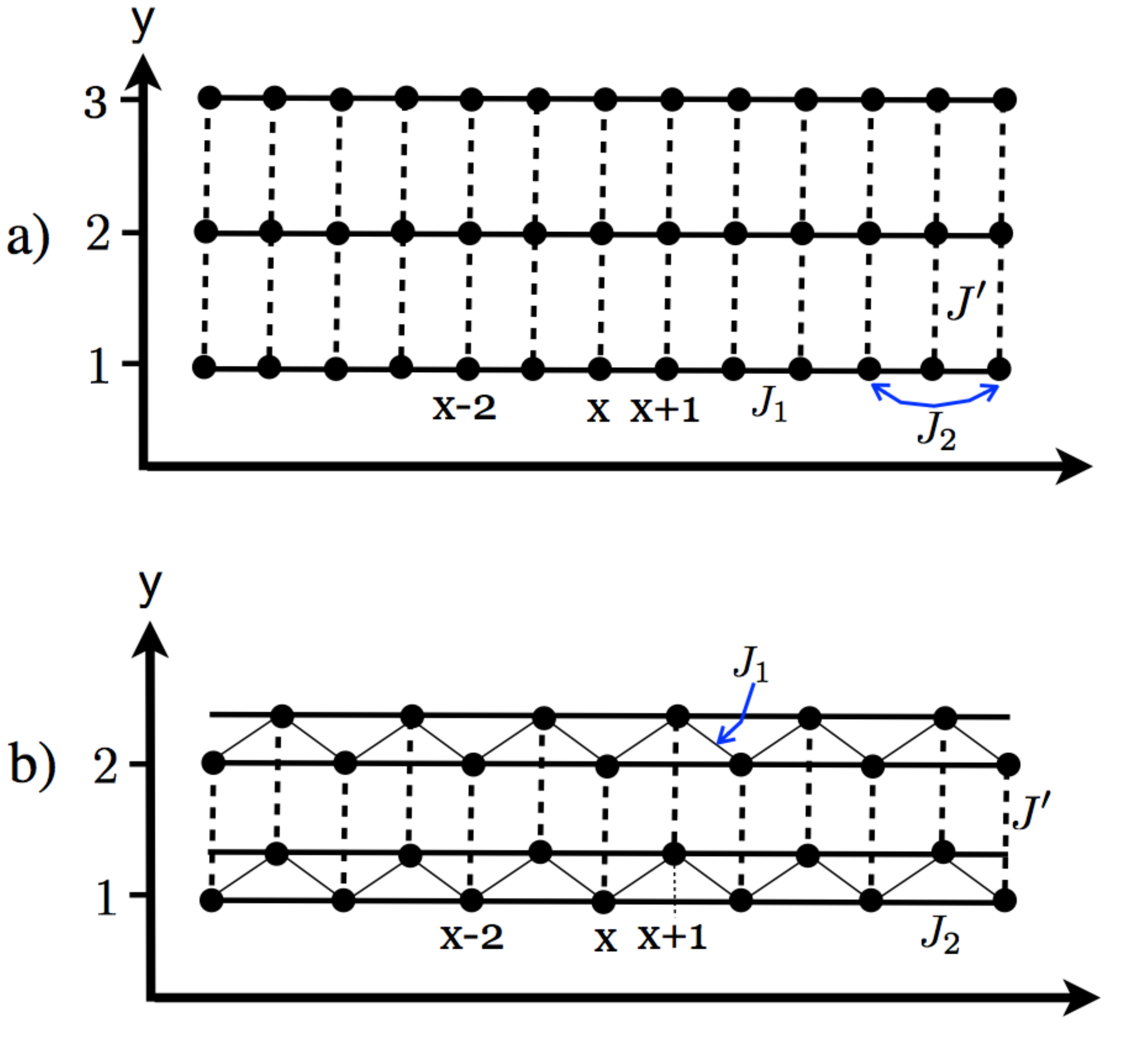}%{DiagramS2a}
\caption{(a) Weakly coupled nematic chains with ferromagnetic $J_1$ (solid lines) and antiferromagnetic next-nearest $J_2$, coupled
by an inter-chain exchange $J'$. (b) Same system viewed as a set of weakly coupled zig-zag chains.}
\label{fig:nematic-chains}
\end{figure}

Note, however, that a true $U(1)$ symmetry breaking is not possible in a single chain, where instead a critical Luttinger state with
algebraically decaying nematic correlations is established \cite{Hikihara2008}. To obtains a true 
{\em two-dimensional nematic} phase, one needs to establish a phase coherence between the phases of order parameters $Q_{--}(y)$
of different chains. 
By our superconducting analogy, this requires a Josephson coupling to transfer (hop)
bound two-magnon pairs between nematic chains. 
The corresponding `hopping' term reads $K \sum_{x,y}( Q_{++}(x,y) Q_{--}(x,y+1) + {\text{h.c.}})$.
Microscopically, such an interaction represents a four-spin coupling, which is not expected
to be particularly large in a good Mott insulator with a large charge gap, such as LiCuVO$_4$. 
However, even if $K$ is absent microscopically, it will be generated perturbatively from the usual
inter-chain spin exchange $J' \sum_{x,y} (S^+_{x,y} S^-_{x,y+1} + {\text{h.c.}})$, 
which plays the role of a single-particle tunneling process in the superconducting analogy.
(Observe that expectation value of this interaction in the chain nematic ground state is zero --
adding or removing of a single magnon to the `superconducting magnon' ground state is forbidden
at energies below the single magnon gap.)
The pair-tunneling generated by fluctuations is estimated to be of the order $K \sim (J')^2/J_1 \ll J'$. 

At the same time, $S^z-S^z$ interaction between chains does not suffer from a similar `low-energy suppression'. This is because 
$S^z$ is simply proportional to a number of magnon pairs, $n_{\rm pair}(x,y) = 2 n(x,y)$, which is just twice the magnon number $n(x,y)$. Hence
$S^z(x,y) = 1/2 - 2 n(x,y)$ differs only a by coefficient $2$ from its usual expression in terms of magnon density.

We thus have a situation where the strength of interchain density-density ($S^z-S^z$) interaction, which is determined by the original
interchain $J'$, is much stronger than that for the fluctuation-generated Josephson interaction $K \sim (J')^2/J_1$. In addition, 
more technical analysis of the scaling dimensions of the corresponding operators shows\cite{Starykh2013}, that the two competing
interactions are characterized by (almost) the same scaling dimension (approximately equal to $1$ for $h \approx h_{\rm sat}$) 
which makes them both strongly relevant in the renormalization group sense. 
Given an inequality $J' \gg (J')^2/J_1$, which selects interchain $S^z-S^z$ interaction as the strongest one, we end up with
a {\em two-dimensional collinear SDW phase} build out of nematic spin chains \cite{sato2013,Starykh2013}. 
This conclusion holds for all $h$ except for the
immediate vicinity of the saturation field $h_{\rm sat}^{(2)}$. There a separate fully two-dimensional 
BEC analysis is required, due to the vanishing of spin velocity at the saturation field, and the result is a true 
{\em two-dimensional nematic phase} in the narrow field range $h_{\rm sat}^{(1)} \lesssim h \leq h_{\rm sat}^{(2)}$ \cite{mzh2010,sato2013,Starykh2013}.
A useful analogy to this  competition is provided by models
of striped superconductors, where the competition is between the superconducting order (a magnetic analogue of which is the spin nematic)
and the charge-density wave order (a magnetic analogue of which is the collinear SDW), see Ref.\onlinecite{Jaefari2010} and references therein.

To summarize, weak inter-chain interaction $J'$ between $J_1-J_2$ spin chains with strong nematic spin correlations 
actually stabilizes a two-dimensional SDW phase as the ground state in a wide range of magnetization. 
This state preserves $U(1)$ symmetry of spin rotations
and is characterized by short-ranged transverse spin correlations, similar to a nematic state.

\subsubsection{Spin-current nematic state at the 1/3-magnetization plateau}

The discussion in the previous Subsection was focused on the systems with ferromagnetic exchange ($J_1 < 0$) on some of the
bonds - as described there, negative exchange needed in order to obtain an {\em attractive} interaction between magnons.

Superconducting analogy, extensively used above, forces one to ask, by analogy with superconducting states of repulsive 
fermion systems (such as, for example, pnictide superconductors or high-temperature cuprate ones), 
if it is possible to realize a spin-nematic in a spin system with only antiferromagnetic 
(that is, repulsive) exchange interactions between magnons. To the best of our knowledge, the first example of such a state is provided by the
spin-current state described in part ({\bf e}) of the Section~\ref{sec:anisotropic-triangular}.
Being nematic, this state is characterized by a spin current long-range order and the absence of 
the magnetic long-range order in the transverse to the magnetic field direction \cite{Chubukov2013}.

A very similar state, named {\em chiral Mott insulator}, was recently discovered in the variational wave function study of a two-dimensional
system of interacting bosons on frustrated triangular lattice \cite{Zaletel2013} as well as in a one-dimensional system of bosons on frustrated ladder
\cite{Dhar2012,Dhar2013}. In both cases, a chiral Mott insulator is an intermediate phase, which separates 
the usual Mott insulator state (which is a boson's analogue of the UUD state) from the superfluid one (which is an analogue of the cone state).
As in Figure~\ref{fig:diag-S}, it intervenes between the states with distinct broken symmetries ($Z_3$ and $O(2)\times Z_2$ in our case),
and gives rise to two continuous transitions instead of a single discontinuous one.

\subsection{Magnetization plateaus in itinerant electron systems}

Up-up-down magnetization plateaus, found in the triangular geometry, are of classical nature. 
Over years, several interesting suggestions of non-classical (liquid-like) magnetization plateaux have been put forward
\cite{Misguich2001,Hida2005,Alicea2007,Tanaka2009,Takigawa2011,Parameswaran2013}, but so far not observed in experiments or numerical simulations. 
Very recently, however, two numerical studies \cite{Nishimoto2013,Capponi2013} of the spin-1/2 kagom\'e antiferromagnet have observed 
non-classical magnetization plateaus at $M/M_{\rm sat} = 1/3, 5/9, 7/9$. These intriguing findings, taken together with
earlier prediction of a {\em collinear spin liquid} at $h = h_{\rm sat}/3$ in the classical kagom\'e antiferromagnet \cite{Zhitomirsky2002},
hint at a very rich magnetization process of the quantum model, the ground state of which at $h=0$ is a $Z_2$ spin liquid! \cite{Yan2011}.

A different point of view on the magnetization plateau was presented recently in Ref.\onlinecite{Hao2013}. The authors asked
if the plateau is possible in an itinerant system of weakly-interacting electrons. The answer to this question is affirmative, as can be understood
from the following consideration. 

\begin{figure}[h]
\centering
\includegraphics[width=7.cm]{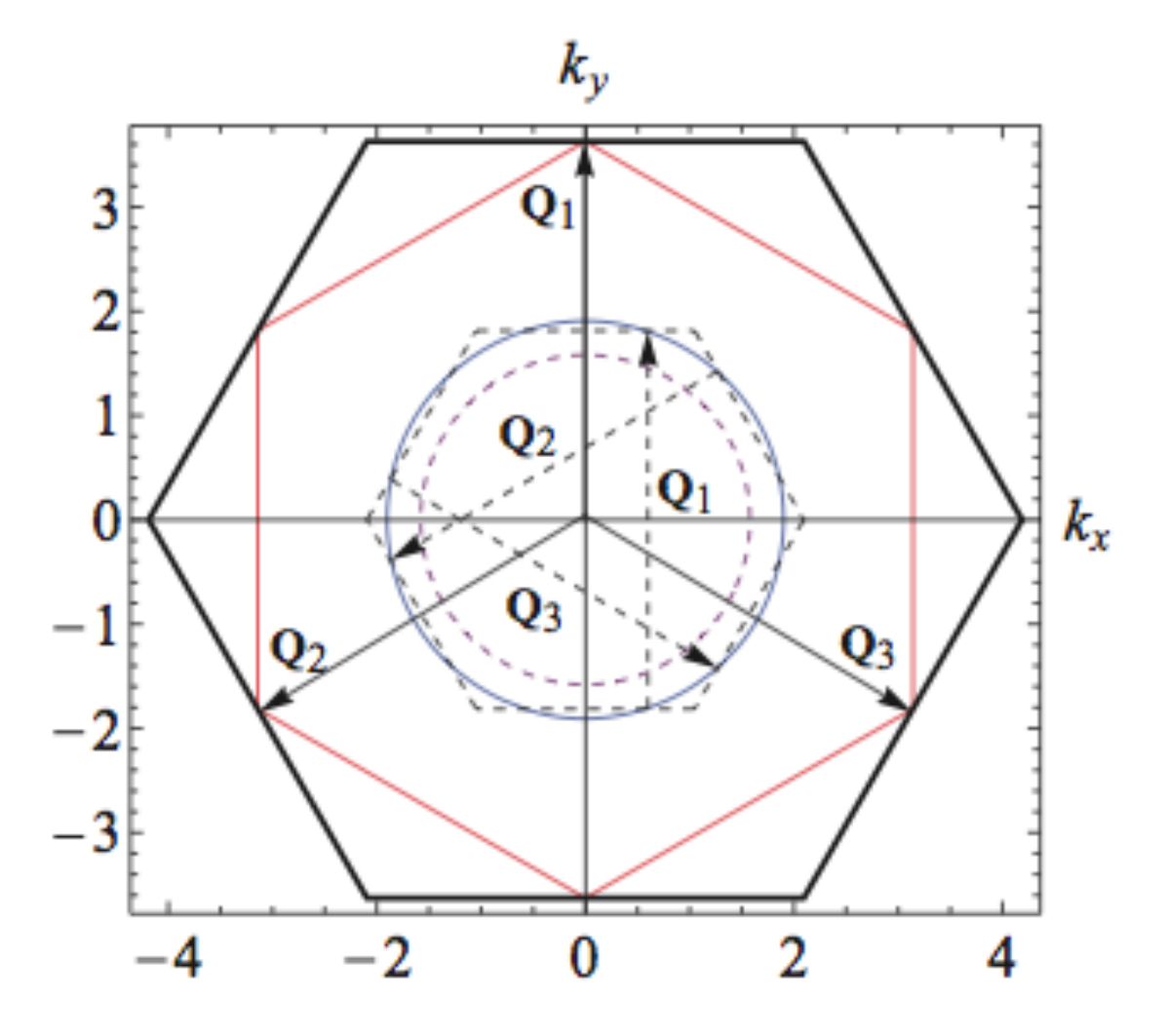}%{DiagramS2a}
\caption{The Fermi surface (red hexagon) of the non-interacting $\sigma = \uparrow$ electrons on a triangular lattice at $n_\uparrow= 3/4$.
Nearly circular Fermi surface of the minority $\sigma = \downarrow$ electrons is shown by blue line. Bold black hexagon represents the Brillouin zone.
Adapted from Zhihao Hao and Oleg A. Starykh, Phys. Rev. B 87, 161109 (2013).}
\label{fig:FS}
\end{figure}

Let us start with a system of non-interacting electrons on a triangular lattice. A magnetic field, applied in-plane in order to avoid complications
due to orbital effects, produces magnetization $M = (n_\uparrow - n_\downarrow)/2$, where densities $n_\sigma$ 
of electrons with spin $\sigma = \uparrow, \downarrow$ are constrained by the total density $n = n_\uparrow + n_\downarrow$.
Consider now special situation with $n_\uparrow= 3/4$, at which the Fermi-surface of $\sigma = \uparrow$ electrons, by virtue of lattice geometry, acquires 
particularly symmetric shape: a hexagon inscribed inside the Brillouin zone hexagon, see Figure~\ref{fig:FS}.

Points where $\sigma = \uparrow$ Fermi-surface touches the Brillouin zone (denoted by vectors $\pm {\bf Q}_j$ with $j=1,2,3$ in Fig.\ref{fig:FS})
are the van Hove points, at which Fermi-velocity vanishes and electron dispersion becomes quadratic. They are characterized by the
logarithmically divergent density of states. In addition, being a hexagon, $\sigma = \uparrow$  Fermi-surface is perfectly nested.
As a result, static susceptibility $\chi_\uparrow({\bf k})$ of spin-up electrons is strongly divergent, as
$\log^2(|{\bf k} - {\bf Q}_j|)$, for wave vectors ${\bf k} \approx {\bf Q}_j$.

Given this highly susceptible spin-$\uparrow$ Fermi surface, it is not surprising that a weak interaction between electrons, 
either in the form of a direct density-density interaction
$V n_{\bf r} n_{{\bf r} + \delta_j}$ between electrons on, e.g., nearest sites, or in the form of a local Hubbard interaction 
$U n_{{\bf r},\uparrow} n_{{\bf r},\downarrow}$ between the majority and minority particles, drives spin-$\uparrow$ electrons
into a gapped correlated state - the charge density wave (CDW) state \cite{Hao2013} with fully gapped Fermi surface.
Moreover, CDW ordering wave vectors ${\bf Q}_j$
are commensurate with the lattice, leading to a commensurate CDW for the spin-$\uparrow$ electrons \cite{Martin2008}.

Minority spin-$\downarrow$ electrons experience position-dependent effective field $U \langle n_{{\bf r},\uparrow}\rangle$ and form
CDW as well. Depending on whether or not vectors ${\bf Q}_j$ span the spin-down Fermi-surface
(this depends on $n_\downarrow$ density), the Fermi-surface of $\sigma = \downarrow$ electrons may or may not 
experience reconstruction. However, being not nested, it is guaranteed to retain at least some parts of the critical Fermi-surface.

The resulting state is a co-existence of a charge- and collinear spin-density waves, together with critical $\sigma = \downarrow$
Fermi surface. Since the energy cost of promoting a spin-$\downarrow$ electron to a spin-$\uparrow$ state is finite (and given by the
gap on the spin-$\uparrow$ Fermi surface), the resulting state realizes {\em fractional} magnetization plateau, the magnetization
of which is determined by the total density $n$ via $M = (3/2 - n)/2$. At half-filling, $n=1$, the plateau is at $1/2$ of the total magnetization,
but  for $n\neq 1$ it takes a fractional value. Amazingly, the obtained state is also a {\em half-metal} \cite{Katsnelson2008} - the only conducting band is
that of (not gapped) minority spin-$\downarrow$ electrons.

Theoretical analysis sketched here bears strong similarities with recent proposals 
\cite{Martin2008,Nandkishore2012,Nandkishore2012a,Kiesel2012,Chern2012} of collinear and chiral spin-density wave 
(SDW) and superconducting states of itinerant electrons on a honeycomb lattice in the vicinity of electron filling factors 3/8 and 5/8 at zero magnetization.
Our analysis shows that even simple square lattice may host similar half-metallic magnetization plateau state, see supplement to \cite{Hao2013}.
Similar to the case of a magnetic insulator, described in the previous Sections, external magnetic field sets the direction of the collinear CDW/SDW state. 
The resulting half-metallic state only breaks the discrete translational symmetry of the lattice, resulting in fully gapped excitations, 
and remains stable to fluctuations of the order parameter about its mean-field value.
In addition to standard solid state settings, the described phenomenon may also be observed in experiments  on cold atoms,
where desired high degree of polarization can be easily achieved \cite{Zwierlein2006}. It appears that, in addition to the 
half-metallic state, the system may also support p-wave superconductivity - a competition between these phases
may be efficiently studied with the help of functional renormalization group \cite{Platt2013}.

\section{Experiments}

Much of the current theoretical interest in quantum antiferromagnetism comes from the amazing experimental progress 
in this area during the last decade. The number of interesting materials is too large to review here, and for this reason
we focus on a smaller sub-set of recently synthesized quantum spin-1/2 antiferromagnets, which realize some of quantum states discussed above.

One of the best known among this new generation of materials is Cs$_2$CuCl$_4$, extensively studied by Coldea and collaborators 
in a series of neutron scattering experiments \cite{Coldea2001,Coldea2002,Coldea2003} and by others via NMR \cite{Vachon2011} and, more recently, ESR 
\cite{povarov2011,Smirnov2012,Zvyagin2013,Fayzullin2013} experiments. This spin-1/2 material represents a realization of a  deformed
triangular lattice with $J'/J = 0.34$ \cite{Coldea2002} and significant DM interactions on chain and inter-chain (zig-zag) bonds, 
connecting neighboring spins \cite{Coldea2002,starykh2010extreme}. Inelastic neutron scattering experiments have revealed unusually strong multi-particle continuum, the origin
of which has sparked intense theoretical debate \cite{Bocquet2001,Alicea2005,Isakov2005,Veillette2005,Veillette2005a,Zheng2006a,Dalidovich2006,Kohno2007}. 
The current consensus is that Cs$_2$CuCl$_4$ is best understood
as a weakly-ordered quasi-one-dimensional antiferromagnet, whose spin excitations smoothly interpolate from fractionalized spin-1/2
spinons of one-dimensional chain at high- and intermediate energies to spin waves at lowest energy ($\ll J'$) \cite{Kohno2007}. 
Although weak, residual inter-plane and DM interactions play the dominant role in the magnetization process of this material. The resulting
$B-T$ phase diagram is rather complex and highly anisotropic \cite{tokiwa2006}, and does not contain a magnetization plateau. However it is worth
mentioning that this was perhaps the first spin-1/2 material, a magnetic response of which featured a SDW-like phase 
ordering wave vector, which scales {\em linearly} with magnetic field in an about 1 Tesla wide interval 
(denoted as phase ``S" in \cite{Coldea2001} and as phase ``E" in \cite{tokiwa2006}). While still not well understood, this experimental 
observation have provided valuable hint to quasi-1d approach based on viewing Cs$_2$CuCl$_4$ as a collection of 
weakly coupled spin chains \cite{Kohno2007}.

\subsection{Magnetization plateau}

Robust $1/3$ magnetization plateau -- the first of its kind among triangular spin-1/2 antiferromagnets -- 
is present in Cs$_2$CuBr$_4$, which has the same crystal structure as Cs$_2$CuCl$_4$,
but is less deformed, $J'/J \approx 0.7 $, and is more two-dimensional than the chloride-based material.

The observed plateau, which is about 1 Tesla wide ($h_{c1} =13.1$T and $h_{c2} = 14.4$T) \cite{Ono2003,Ono2004,Ono2005}, 
is clearly visible in both magnetization  and elastic neutron scattering measurements \cite{Ono2004,Ono2005}, which determined the UUD spin
structure on the plateau. The observation of the magnetization plateau has generated a lot of experimental activity. The quantum origin of the plateau visibly manifests itself
via essentially temperature-independent plateau's critical fields $h_{c1,2}(T) \approx h_{c1,2}(T=0)$, as found in the thermodynamic
study \cite{Tsujii2007}. This behavior should be contrasted with the phase diagram of the spin-$5/2$ antiferromagnet 
RbFe(MoO$_4$)$_2$ \cite{Svistov2006,Smirnov2007,White2013}, where the critical field $h_{c1}(T)$ does show strong downward shift with $T$.
(Recall that in the classical model, Figure~\ref{fig:diagram}, the UUD phase collapses to a single point at $T=0$.) 

\begin{figure}[h]
\centering
\includegraphics[width=8.cm]{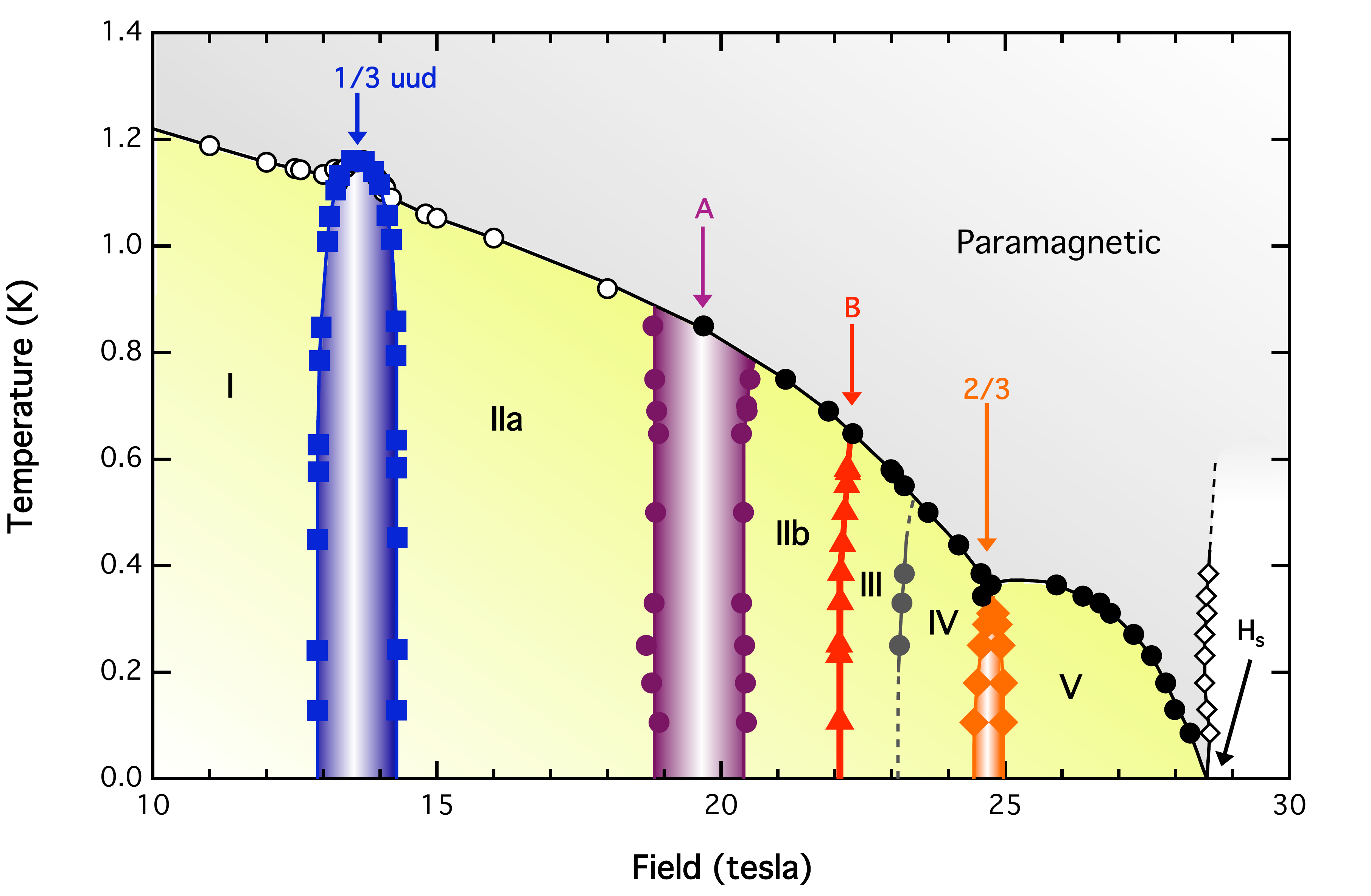}%{DiagramS2a}
\caption{Magnetic phase diagram of Cs$_2$CuBr$_4$, as deduced from the magnetocaloric-effect data taken at various temperatures. 
Circles indicate second-order phase boundaries, whereas other symbols except the open diamonds indicate first-order boundaries.
Adapted from Fortune {\it et al.}, Phys. Rev. Lett. 102, 257201 (2009).}
\label{fig:cs2cubr4}
\end{figure}

Commensurate up-up-down spin structure of Cs$_2$CuBr$_4$ is also supported
by NMR measurements \cite{Fujii2004,Fujii2007} which in addition finds that transitions between the commensurate 
plateau and adjacent to it incommensurate phases are discontinuous (first-order). 
Extensive magnetocaloric effect and magnetic-torque experiments \cite{Fortune2009} have uncovered surprising 
cascade of field-induced phase transitions in the interval $10 - 30$ T. The most striking feature of the emerging complex phase 
diagram is that it appears to contain up to 9 different magnetic phases -- in stark contrast with the `minimal' theoretical model diagram
in Figure~\ref{fig:diagram} which contains just 3 phases! This, as well as strong sensitivity of the magnetization curve to the
direction of the external magnetic field with respect to crystal axis, strongly suggest that the difference in the phase diagrams 
has to do with spatial ($J' \neq J$) and spin-space (asymmetric DM interaction) anisotropies present in Cs$_2$CuBr$_4$.
Large-$S$ and classical Monte Carlo studies \cite{griset2011deformed} do find the appearance of new incommensurate
phases in the phase diagram, in qualitative agreement with the large-$S$ diagram of Figure~\ref{fig:diag-S} (note that the latter
does not account for the DM interaction which significantly complicates the overall picture \cite{griset2011deformed}).

Perhaps the most puzzling of the ``six additional" phases is a
narrow region at about $B=23$T, where $dM/dB$ exhibits sharp double peak structure, interpreted in \cite{Ono2005,Ono2011}
as a novel magnetization plateau at $M/M_{\rm sat} = 2/3$. Such a new $2/3$-magnetization plateau was observed in an exact diagonalization study
of spatially anisotropic spin-1/2 model \cite{Miyahara2006} but was not seen in more recent variational wave function
 \cite{tay2010variational} and DMRG \cite{chen2013}, as well as in analytical 
 large-$S$ \cite{alicea2009quantum,griset2011deformed} studies.

Nearly isotropic, $J'/J \approx 1$, antiferromagnet Ba$_3$CoSb$_2$O$_9$ is believed to provide an `ideal' realization
of the spin-1/2 antiferromagnet on a uniform triangular lattice \cite{Susuki2013,Shirata2012}. And, indeed, its experimental phase diagram
is in close correspondence with $J'=J$ `cut in Figure~\ref{fig:diag-S} 
(along $\delta=0$ line) and Figure~\ref{fig:diag1/2} (along $R=0$ line): it has $120^\circ$ spin structure at zero field, 
coplanar Y state at low fields, the 1/3 magnetization plateau in the $h_{c1}/h_{\rm sat} = 0.3 \leq h/h_{\rm sat} \leq h_{c2}/h_{\rm sat} = 0.47$ interval,
and coplanar V state (denoted as $2:1$ state in \cite{Susuki2013}) at higher fields. 

A new element of the study \cite{Susuki2013,Shirata2012} is the appearance 
of weak anomaly in $dM/dB$ at about $M/M_{\rm sat} = 3/5$, which was interpreted as a quantum phase transition
from the V phase to another coplanar phase - inverted Y (state `e' in Figure~\ref{fig:states}). Near the saturation field these two phases are very close in energy, 
the difference appears only in the 6th order in condensate amplitude \cite{Nikuni1995}. Such a transition
can be driven by sufficiently strong easy-plane anisotropy \cite{Yamamoto2013} as well as anisotropic DM interaction \cite{griset2011deformed}.

Perhaps, the more relevant to Ba$_3$CoSb$_2$O$_9$ is another possibility - that a transition is driven by the interlayer interaction.
Ref.~\onlinecite{Gekht1997} has shown that weak inter-plane antiferromagnetic exchange interaction
causes transition from the uniform V phase to the {\em staggered V} phase. The latter is described by the same Eq.\eqref{eq:V} but with a
$z$-dependent phase, $\varphi_z = \tilde{\varphi} + \pi z$ (here, $z$ is the integer coordinate of the triangular layer and $\tilde{\varphi}$ is an
overall constant phase), leading to the doubling of the period of the magnetic structure along the  direction normal to the layer. 
It is easy to see that such a state actually gains energy from the antiferromagnetic interlayer exchange $J''$,
while preserving the optimal in-plane configuration in every layer.
Such a transition, denoted as HFC1-HFC2 transition, was also observed in recent semi-classical Monte Carlo simulations \cite{Koutroulakis2013}.
This development suggest that a mysterious `2/3-plateau' of Cs$_2$CuBr$_4$, mentioned above,  may too be related to
a transition between the lower-field uniform and higher-field staggered versions of the commensurate V phase.

\subsection{SDW and spin nematic phases}
\label{sec:sdw-exp}

A collinear SDW order has been observed in 
spin-1/2 Ising-like antiferromagnet BaCo$_2$V$_2$O$_8$.
Experimental confirmations of this comes from specific heat \cite{Kimura2008} and neutron diffraction \cite{Kimura2008a} measurements.
The latter one is particularly important as it proofs the linear scaling of the SDW ordering wave vector with the magnetization,
$k_{\rm sdw} = \pi(1-2M)$, predicted in \cite{Suzuki2007}.
Subsequent NMR \cite{Klanjsek2012}, ultrasound \cite{Yamaguchi2011}, and neutron scattering \cite{Canevet2013} 
experiments have refined the phase diagram and even proposed the existence of two different SDW phases \cite{Klanjsek2012}
stabilized by competing interchain interactions.

Most recently, spin-1/2 magnetic insulator LiCuVO$_4$ has emerged \cite{enderle2005,Nishimoto2012} as a promising candidate to realize both a high-field 
spin nematic phase, right below the two-magnon saturation field, which is about $45$ T high, and an incommensurate collinear 
SDW phase at lower fields, extending from about $40$T down to about $10$ T. At yet lower magnetic field, the material realizes more conventional vector chiral
(umbrella) state which can be stabilized by a moderate easy-plane anisotropy of exchange interactions \cite{Heidrich-Meisner2009}
(which does not affect the high field physics discussed here).

This last material seems to nicely realize theoretical scenario 
outlined in Section~\ref{sec:nematic-chains}:  spin-nematic chains \cite{kolezhuk05,Vekua2007} form a 
two-dimensional nematic phase only in the immediate vicinity of the saturation field \cite{Svistov2011}. At fields below that rather
narrow interval, the ground state is an incommensurate longitudinal SDW state. Evidence for the latter includes detailed
studies of NMR line shape \cite{buttgen2007,Buttgen2010,svistov2012,Nawa2013} and neutron scattering \cite{masuda2011,mourigal2012}.
It is worth adding here that quasi-one-dimensional nature of this material is evident from the very pronounced multi-spinon continuum,
observed at $h=0$ in inelastic neutron scattering studies \cite{Enderle2010}.

\subsection{Weak Mott insulators: Hubbard model on anisotropic triangular lattice}

Given that, quite generally, Heisenberg Hamiltonian can be viewed as a strong-coupling (large $U/t$) limit of the Hubbard model, 
it is natural to consider the fate of the Hubbard $t - t' - U$ model on (spatially anisotropic, in general) triangular lattice.

As a matter of fact, this very problem is of immediate relevance to intriguing experiments on organic Mott insulators 
of X[Pd(dmit)$_2$]$_2$ and $\kappa$-(ET)$_2$Z families. 
Recent experimental \cite{Kanoda2011,Yamashita2012} and theoretical \cite{Lee2006,Sachdev2009,Balents2010,Powell2011} reviews describe key 
relevant to these materials issues, and we direct interested readers to these publications. 

\begin{figure}[h]
\centering
\includegraphics[width=9.cm]{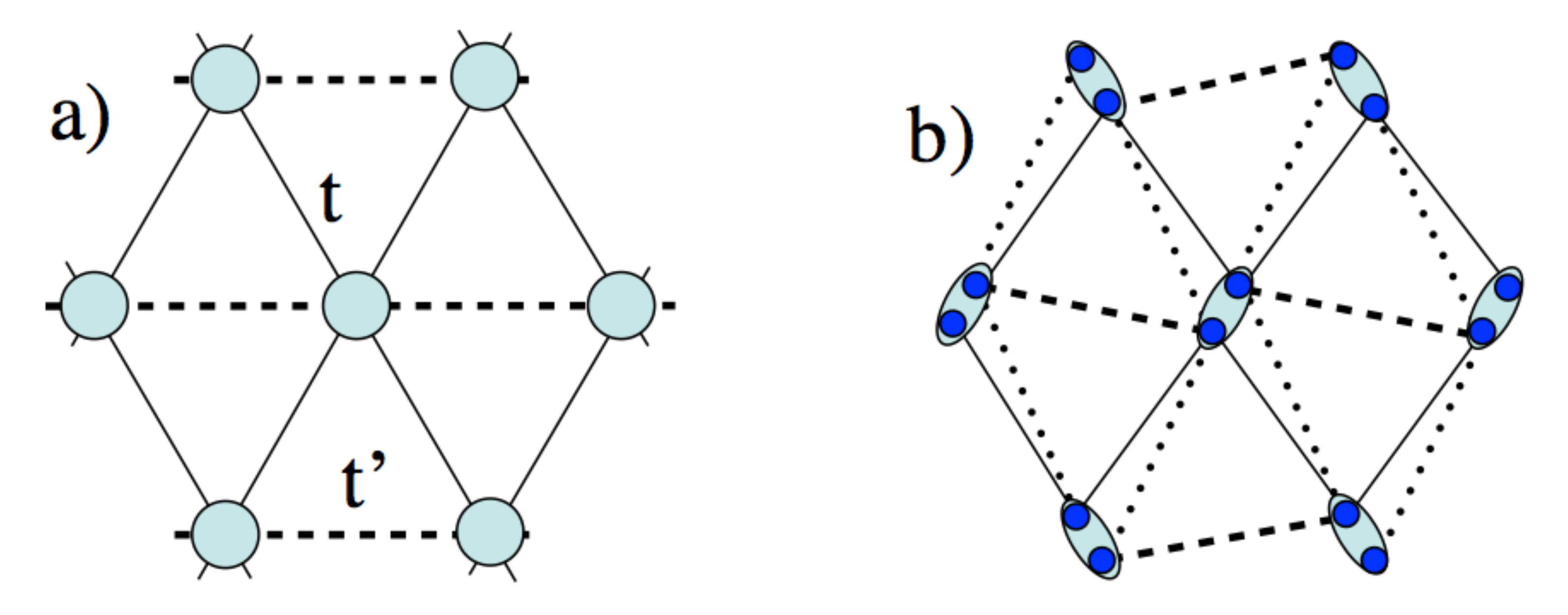}%{DiagramS2a}
\caption{(a) Spatially anisotropic single-band Hubbard model for organic Mott insulators, after Refs.\onlinecite{Kanoda2011,Hiroi2011}. Note that 
$t'/t < 1$ here implies $J/J' < 1$ in Figure~\ref{fig:lattice}: $J'$ of Fig.~\ref{fig:lattice} actually lives on $t$ bonds of the Hubbard model here.
(b) Geometry of an extended two-band model, after Ref.\onlinecite{Hotta2010}. The 'sites', shown by filled dots inside 
ovals representing dimers, are now two-molecule dimers. Two sites from the same dimer are connected by the intra-dimer tunneling amplitude $t_d$, and in the limit 
$t_d \gg {\text{'inter-dimer tunnelings'}}$ the model reduces to that in (a). Note also the appearance of additional hopping amplitudes 
(dotted line) connecting 'more distant' sites of neighboring dimers.}
\label{fig:ET}
\end{figure}

One of the main unresolved issues in this field is that of a proper minimal model that captures all relevant degrees of freedom.
Highly successful initial proposal \cite{Kino1995} models the system as a simple half-filled Hubbard model on a spatially anisotropic triangular lattice,
as sketched in Figure~\ref{fig:ET}. Every `site' of this lattice in fact represents closely bound dimer, made of two ET molecules \cite{Kino1995},
and occupied by single electron (or hole).
Spatial anisotropy shows up via different hopping integrals $t, t'$. Within the standard large-$U$ description,
anisotropy of hopping $t'/t$ directly translates into that of exchange interactions on different bonds, $J'/J \sim (t'/t)^2$.
Ironically, most of the studied materials fall onto $t' < t$ side \cite{Kanoda2011,Hiroi2011} of the diagram, which happens to be opposite to $J > J'$
limit of spatially anisotropic Heisenberg model, to which this review is devoted.

This description has generated a large number of interesting studies, the full list of which is beyond the scope of this review.
One of the main outcomes of these studies is the establishment of approximate $t'/t - U$ phase diagram (see for example Figure~6 of \cite{Tocchio2013})
which harbors metallic phase (for $U/t \lesssim 10$ or less, depending on $t'/t$) and various insulating magnetic phases,
which include both the standard N\'eel and non-coplanar spiral phases as well as quantum-disordered spin-liquid state (for $t'/t \approx 0.9$ and $U/t \gtrsim 12$).

Thinking in terms of effective spin-only model, it is important to realize that for not too large $U/t$, the standard Heisenberg model must be amended 
with {\em ring-exchange} terms involving four (or more) long spin loops \cite{Misguich1998,Motrunich2005}. This addition dramatically affects the regime of intermediate $U/t$ 
by stabilizing an insulating spin-liquid ground state \cite{Motrunich2005,Yang2010}. The nature of the emerging spin-liquid is subject of intense 
on-going investigations, with proposals ranging from $Z_2$ liquid \cite{Xu2009,Barkeshli2013} to spin Bose-metal \cite{Sheng2009},
to spin-liquid with quadratic band touching \cite{Mishmash2013}.

Recently, however, this appealing spin-only picture of the organic Mott insulators has been challenged by the experimental discovery of anomalous
response of dielectric constant \cite{Abdel-Jawad2010} and lattice expansion coefficient \cite{Manna2010} at low temperature. 
This finding imply that charge degrees of freedom, assumed frozen in the spin-only description, are actually present in the material and have to be
accounted for in theoretical modeling. Several subsequent papers \cite{Hotta2010,Naka2010,Gomes2013} have identified dimer units of the triangular lattice,
viewed as sites in Figure~\ref{fig:ET}, as the most likely place where charge dynamics persists down to lowest temperatures. To describe these
internal states of the two-molecule dimers, one need to go back to a two-band extended Hubbard model description \cite{Seo2004}.
Taking the strong-coupling of such a model, one derives \cite{Hotta2010} a coupled dynamics of interacting spins and {\em electric dipoles} 
on the triangular lattice. In turns out that sufficiently strong inter-dimer Coulomb interaction stabilizes charge-ordered state (dipolar solid) and
suppresses spin ordering via non-trivial modification of exchange interactions $J, J'$.

Clearly many more studies, both experimental and theoretical, are required in order to elucidate the physics behind apparent spin-liquid 
behavior of organic Mott insulators.

{\bf In place of conclusion} we just state the obvious:
despite many years of investigations, quantum magnets on triangular lattices continue to surprise us. There are no doubts that future studies 
of new materials and models, inspired by them, will bring out new quantum states and phenomena.

\begin{acknowledgments}

I am grateful to my
friends and coauthors - Leon Balents, Andrey Chubukov, Jason Alicea and Zhihao Hao -
for fruitful collaborations and countless insightful discussions that provide the foundation of this review.
I thank Sasha Abanov, Hosho Katsura, Ru Chen, Hyejin Ju, Hong-Chen Jiang, Christian Griset, and Shane Head for their crucial contributions 
to joint investigations related to the topics discussed here. 
Discussions of experiments with Collin Broholm, Radu Coldea, Martin Mourigal, Masashi Takigawa, Leonid Svistov, Alexander Smirnov and Yasu Takano are greatly appreciated as well. 
I have benefited extensively from conversations with Cristian Batista, Misha Raikh, Dima Pesin, Eugene Mishchenko and Oleg Tchernyhyov.
Many thanks to Luis Seabra, Nic Shannon, Alexander Smirnov and Yasu Takano for permissions to reproduce figures from their papers in this review.
Special thanks to Andrey Chubukov for the critical reading of the manuscript and invaluable comments.
This work is supported by the National Science Foundation through grant DMR-12-06774.
\end{acknowledgments}

\bibliographystyle{nsf_bib_style}
\bibliography{ropp-arxiv.bib}
\end{document}